\documentclass[submission,copyright,creativecommons]{eptcs}
\providecommand{\event}{Thiemann Festschrift} 

\usepackage{iftex}

\ifpdf
  \usepackage{underscore}         
  \usepackage[T1]{fontenc}        
\else
  \usepackage{breakurl}           
\fi

\usepackage{xspace}

\usepackage{tikz-cd}
\usepackage{fancyvrb}
\usepackage{latex/agda}
\usepackage{newunicodechar}
\usepackage{etoolbox}
\undef{\hbar}
\usepackage{oz}

\renewcommand{\AgdaCodeStyle}{\small}
\newcommand{\myrule}{\rule{0.25\textwidth}{0.5pt}}

\usepackage{inconsolata}

\newcommand{\lambdabar}{{\mkern0.75mu\mathchar '26\mkern -9.75mu\lambda}}
\usepackage[utf8]{inputenc}
\usepackage{newunicodechar}
\newunicodechar{λ}{\ensuremath{\mathnormal\lambda}}
\newunicodechar{∀}{\ensuremath{\mathnormal\forall}}
\newunicodechar{₀}{\ensuremath{\texttt{\tiny 0}}}
\newunicodechar{₁}{\ensuremath{\texttt{\tiny 1}}}
\newunicodechar{₂}{\ensuremath{\texttt{tiny 2}}}
\newunicodechar{≡}{\ensuremath{\mathnormal\equiv}}
\newunicodechar{σ}{\ensuremath{\mathnormal\sigma}}
\newunicodechar{∣}{\ensuremath{\mathnormal\vert}}
\newunicodechar{∎}{\ensuremath{\mathnormal\blacksquare}}
\newunicodechar{⊢}{\ensuremath{\mathnormal\vdash}}
\newunicodechar{⊨}{\ensuremath{\mathnormal\models}}
\newunicodechar{●}{\ensuremath{\mathnormal\bullet}}
\newunicodechar{ƛ}{\ensuremath{\mathnormal\lambdabar}}
\newunicodechar{▷}{\ensuremath{\mathnormal\vartriangleright}}
\newunicodechar{Γ}{\ensuremath{\mathnormal\Gamma}}
\newunicodechar{Δ}{\ensuremath{\mathnormal\Delta}}
\newunicodechar{Θ}{\ensuremath{\mathnormal\Theta}}
\newunicodechar{Ξ}{\ensuremath{\mathnormal\Xi}}
\newunicodechar{τ}{\ensuremath{\mathnormal\tau}}
\newunicodechar{υ}{\ensuremath{\mathnormal\upsilon}}
\newunicodechar{ℕ}{\ensuremath{\mathnormal\mathbb N}}
\newunicodechar{⇒}{\ensuremath{\mathnormal\Rightarrow}}
\newunicodechar{⇛}{\ensuremath{\mathnormal\Rrightarrow}}
\newunicodechar{∅}{\ensuremath{\mathnormal\emptyset}}
\newunicodechar{△}{\ensuremath{\mathnormal\triangle}}
\newunicodechar{⨾}{\ensuremath{\mathnormal\fcmp}}
\newunicodechar{⨟}{\ensuremath{\mathnormal\fcmp}}
\newunicodechar{β}{\ensuremath{\mathnormal\beta}}
\newunicodechar{η}{\ensuremath{\mathnormal\eta}}

\usepackage{hyperref}
\hypersetup{colorlinks=true, linkcolor=black, citecolor=black, filecolor=black, urlcolor=black}

\tikzcdset{every label/.append style = {font = \normalsize}}

\title{Explicit Weakening}
\author{Philip Wadler
\institute{University of Edinburgh}
\email{wadler@inf.ed.ac.uk}
}
\def\titlerunning{Explicit Weakening}
\def\authorrunning{Philip Wadler}

\newcommand{\pause}{\vspace{2ex}}
\begin{document}
\maketitle

\begin{abstract}
I present a novel formulation of substitution, where facts
about substitution that previously required tens or hundreds of lines
to justify in a proof assistant now follow immediately---they can be
justified by writing the four letters "refl".
The paper is an executable literate Agda script,
and source of the paper is available as an artifact in the file
\begin{center}
Weaken.lagda.md
\end{center}
\pause
Not all consequences of the pandemic have been awful. For the last
three years, I've had the great pleasure of meeting with Peter
Thiemann and Jeremy Siek for a couple of hours every week, via Zoom,
exploring topics including core calculi, gradual typing, and
formalisation in Agda. The work reported here arose from those
discussions, and is dedicated to Peter on the occasion of his 60th
birthday.
\end{abstract}

\hypertarget{introduction}{%
\section{Introduction}\label{introduction}}

Every user of a proof assistant has suffered the plague of lemmas:
sometimes a fact that requires zero lines of justification for a
pen-and-paper proof may require tens or hundreds of lines of
justification when using a proof assistant.

Properties of substitution often give rise to such an inundation of
lemmas. To give one example---the one which motivated this work---I
recall a proof formalised in Agda of the gradual guarantee for a
simply-typed gradual cast calculus. The calculus uses lambda terms with
de Bruijn indices, and the proof depends crucially on the following
equivalence.

\begin{verbatim}
(N ↑) [ M ]₀  ≡  N                                                             (*)
\end{verbatim}

Here \texttt{N\ {[}\ M\ {]}₀} substitutes term \texttt{M} for de Bruijn
index zero in term \texttt{N}, while \texttt{N\ ↑} increments by one
every free de Bruijn index in term \texttt{N}. Hence \texttt{N\ ↑} will
not contain de Bruijn index zero, and the equation appears obvious.
However, my formal proof required eleven lemmas and 96 non-comment lines
of code to establish the above fact.

Here I present a novel formulation of substitution where equation (*) is
obvious to the proof assistant as well as to the person conducting the
proof: it holds by definitional equality. Further, many other equations
that one needs also hold by definition and much of the remaining work
can be handled by automatically applied rewrites. As a result, the facts
about substitution that we need can be proved trivially. In my proof
assistant of choice, Agda, one needs to write just four letters,
\texttt{refl}, denoting proof by reflexivity.

Since a single example may fail to convince, I give two more.

First, in the proof of the gradual guarantee mentioned earlier, it is
not just the above equation but \emph{every} property of substitution in
the proof that is rendered trivial by the new formulation given here.

Second, consider an example from the textbook \emph{Programming Language
Foundations} in Agda, by Kokke, Siek, and Wadler \cite{PLFA,PLFA-paper},
henceforth PLFA. Chapter Substitution is devoted to proving the
following equation, for terms using de Bruijn indexes.

\begin{verbatim}
N [ M ]₀ [ L ]₀ ≡ N [ L ]₁ [ M [ L ]₀ ]₀
\end{verbatim}

Here \texttt{N\ {[}\ M\ {]}₀}, as before, substitutes term \texttt{M}
for de Bruijn index zero in term \texttt{N}, while
\texttt{N\ {[}\ L\ {]}₁} substitutes term \texttt{L} for de Bruijn index
one in term \texttt{N}. The equation states that we can substitute in
either order: we can first substitute \texttt{M} into \texttt{N} and
then substitute \texttt{L} into the result, or we can first substitute
\texttt{L} into \texttt{N} and then substitute \texttt{M} (adjusted by
substituting \texttt{L} into \texttt{M}) into the result. The chapter
takes several hundred lines to achieve the above result, and was
considered so long and tedious that it was relegated to an appendix.
With the new formulation, the result becomes immediate.

\pause

My formulation is inspired by the explicit substitutions of Abadi,
Cardelli, Curien, and Levy \cite{ACCL,ACCL-journal}, henceforth ACCL,
one of the seminal works in the area. ACCL wished to devise a calculus
that could aid in the design of implementations, while we focus on
support for automated reasoning; and ACCL considered applications to
untyped, simply-typed, and polymorphic lambda calculus (System F), while
we focus on simply-typed lambda calculus. A vast body of literature is
devoted to explicit substitution. Helpful surveys include Kesner
\cite{Kesner-2009} and Rose, Bloo, and Lang \cite{Rose-Bloo-Lang-2012}.
Another inspiration for this work is the Autosubst system of Schafer,
Tebbi, and Smolka \cite{Schafer-Tebbi-Smolka-2015}.

Despite the large number of variations that have been considered, I have
not found a formulation that matches the one given here. The closest are
David and Guillaume \cite{David-Guillaume-2001} and Hendricks and van
Oostrom \cite{Hendricks-van-Oostrom-2003} both of whom use an explicit
weakening, but with quite different goals. In particular, both used
named variables and single substitution, whereas de Bruijn variables and
simultaneous substitution are crucial to both ACCL and the approach
taken here.

The formulation given here is couched in terms of de Bruijn indices and
intrinsic types, as first proposed by Altenkirch and Reus
\cite{Altenkirch-Reuss-1999}. Lambda calculus is typically formulated
using named variables and extrinsic typing rules, but when using a proof
assistant it is often more convenient to use de Bruijn indices and
intrinsic typing rules. With named variables the Church numeral two is
written \texttt{λs.λz.(s(sz))}, whereas with de Bruijn indices it is
written \texttt{λλ(1(10))}. In the latter variables names do not appear
at point of binding, and instead each variable is replaced by a count
(starting at zero) of how many binders outward one must step over to
find the one that binds this variable. With extrinsic typing, one first
gives a syntax of pre-terms and then gives rules assigning types to
terms, while with intrinsic typing the syntax of terms and the type
rules are defined together. Reynolds \cite{Reynolds-2000} introduced the
names intrinsic and extrinsic; the distinction between the two is
sometimes referred to as Curry-style (terms exist prior to types) and
Church-style (terms make sense only with their types). A textbook
development in Agda of both the named-variable/extrinsic and the de
Bruijn/intrinsic style can be found in PLFA.

The calculus in ACCL is called \emph{explicit substitution}, or
\texttt{λσ}. Our calculus is called \emph{explicit weakening}, or
\texttt{λ↑}. In ACCL, substitutions are constructed with four operations
(id, shift, cons, compose) and substitution is an explicit operator on
terms. Here, substitutions are constructed with three operations (id,
weaken, cons, where weaken is a special case combining shift with
composition) and weakening is an explicit operator on terms, while
substitution and composition become meta operations. For ACCL it is
important that substitution is explicit, as this is key in using the
calculus to design efficient implementations. Conversely, for us it is
important that only weakening is explicit and that substitution and
composition are meta operations, as this design supports the proof
assistant in automatically simplifying terms---so that equations which
were previously difficult to prove become trivial.

\pause

The remainder of this paper develops the new formulation as a literate
Agda script. Every line of Agda code is included in this paper, and the
source is provided as an artifact. Since the paper is a literate Agda
script, when you see code in colour that means it has been type-checked
by the Agda system, providing assurance that it is correct. The paper is
intended to be accessible to anyone with a passing knowledge of proof
assistants. Additional detail on how to formalise proofs in Agda can be
found in PLFA \cite{PLFA,PLFA-paper}.

\hypertarget{formal-development}{%
\section{Formal development}\label{formal-development}}

\hypertarget{module-and-imports}{%
\subsection{Module and imports}\label{module-and-imports}}

We begin with bookkeeping: an option pragma that enables rewriting, the
module header, and imports from the Agda standard library. The first two
imports are required by the pragma, and the others import equality and
related operations. Agda supports mixfix syntax, so \texttt{\_≡\_} names
infix equality.

\begin{code}%
\>[0]\AgdaSymbol{\{-\#}\AgdaSpace{}%
\AgdaKeyword{OPTIONS}\AgdaSpace{}%
\AgdaPragma{--rewriting}\AgdaSpace{}%
\AgdaSymbol{\#-\}}\<%
\\
\>[0]\AgdaKeyword{module}\AgdaSpace{}%
\AgdaModule{Weaken}\AgdaSpace{}%
\AgdaKeyword{where}\<%
\\
\>[0]\AgdaKeyword{open}\AgdaSpace{}%
\AgdaKeyword{import}\AgdaSpace{}%
\AgdaModule{Agda.Builtin.Equality}\<%
\\
\>[0]\AgdaKeyword{open}\AgdaSpace{}%
\AgdaKeyword{import}\AgdaSpace{}%
\AgdaModule{Agda.Builtin.Equality.Rewrite}\<%
\\
\>[0]\AgdaKeyword{import}\AgdaSpace{}%
\AgdaModule{Relation.Binary.PropositionalEquality}\AgdaSpace{}%
\AgdaSymbol{as}\AgdaSpace{}%
\AgdaModule{Eq}\<%
\\
\>[0]\AgdaKeyword{open}\AgdaSpace{}%
\AgdaModule{Eq}\AgdaSpace{}%
\AgdaKeyword{using}\AgdaSpace{}%
\AgdaSymbol{(}\AgdaOperator{\AgdaDatatype{\AgdaUnderscore{}≡\AgdaUnderscore{}}}\AgdaSymbol{;}\AgdaSpace{}%
\AgdaInductiveConstructor{refl}\AgdaSymbol{;}\AgdaSpace{}%
\AgdaFunction{cong}\AgdaSymbol{;}\AgdaSpace{}%
\AgdaFunction{cong₂}\AgdaSymbol{)}\<%
\\
\>[0]\AgdaKeyword{open}\AgdaSpace{}%
\AgdaModule{Eq.≡-Reasoning}\AgdaSpace{}%
\AgdaKeyword{using}\AgdaSpace{}%
\AgdaSymbol{(}\AgdaOperator{\AgdaFunction{begin\AgdaUnderscore{}}}\AgdaSymbol{;}\AgdaSpace{}%
\AgdaFunction{step-≡-∣}\AgdaSymbol{;}\AgdaSpace{}%
\AgdaOperator{\AgdaFunction{\AgdaUnderscore{}∎}}\AgdaSymbol{)}\<%
\end{code}

(As of agda-stdlib-2.1, one imports \texttt{step-≡-∣} to define the
operator \texttt{\_≡⟨⟩\_} used to display chains of equalities.)

\hypertarget{operator-priorities}{%
\subsection{Operator priorities}\label{operator-priorities}}

We declare in advance binding priorities for infix, prefix, and postfix
operators. A higher priority indicates tighter binding, and letters
\texttt{l} or \texttt{r} indicate left or right associativity.

\begin{code}%
\>[0]\AgdaKeyword{infix}%
\>[8]\AgdaNumber{4}%
\>[11]\AgdaOperator{\AgdaDatatype{\AgdaUnderscore{}⊢\AgdaUnderscore{}}}%
\>[16]\AgdaOperator{\AgdaDatatype{\AgdaUnderscore{}⊨\AgdaUnderscore{}}}\<%
\\
\>[0]\AgdaKeyword{infixl}%
\>[8]\AgdaNumber{5}%
\>[11]\AgdaOperator{\AgdaInductiveConstructor{\AgdaUnderscore{}▷\AgdaUnderscore{}}}%
\>[16]\AgdaOperator{\AgdaFunction{\AgdaUnderscore{}⨾\AgdaUnderscore{}}}\<%
\\
\>[0]\AgdaKeyword{infix}%
\>[8]\AgdaNumber{5}%
\>[11]\AgdaOperator{\AgdaInductiveConstructor{ƛ\AgdaUnderscore{}}}\<%
\\
\>[0]\AgdaKeyword{infixl}%
\>[8]\AgdaNumber{7}%
\>[11]\AgdaOperator{\AgdaInductiveConstructor{\AgdaUnderscore{}·\AgdaUnderscore{}}}\<%
\\
\>[0]\AgdaKeyword{infixr}%
\>[8]\AgdaNumber{7}%
\>[11]\AgdaOperator{\AgdaInductiveConstructor{\AgdaUnderscore{}⇒\AgdaUnderscore{}}}\<%
\\
\>[0]\AgdaKeyword{infix}%
\>[8]\AgdaNumber{8}%
\>[11]\AgdaOperator{\AgdaInductiveConstructor{\AgdaUnderscore{}↑}}\<%
\end{code}

\hypertarget{types}{%
\subsection{Types}\label{types}}

A type is either the natural number type (\texttt{\textasciigrave{}ℕ})
or a function type (\texttt{\_⇒\_}).

\begin{code}%
\>[0]\AgdaKeyword{data}\AgdaSpace{}%
\AgdaDatatype{Type}\AgdaSpace{}%
\AgdaSymbol{:}\AgdaSpace{}%
\AgdaPrimitive{Set}\AgdaSpace{}%
\AgdaKeyword{where}\<%
\\
\>[0][@{}l@{\AgdaIndent{0}}]%
\>[2]\AgdaInductiveConstructor{`ℕ}%
\>[7]\AgdaSymbol{:}%
\>[10]\AgdaDatatype{Type}\<%
\\
\>[2]\AgdaOperator{\AgdaInductiveConstructor{\AgdaUnderscore{}⇒\AgdaUnderscore{}}}%
\>[7]\AgdaSymbol{:}%
\>[10]\AgdaDatatype{Type}\AgdaSpace{}%
\AgdaSymbol{→}\AgdaSpace{}%
\AgdaDatatype{Type}\AgdaSpace{}%
\AgdaSymbol{→}\AgdaSpace{}%
\AgdaDatatype{Type}\<%
\end{code}

We let \texttt{A}, \texttt{B}, \texttt{C} range over types.

\begin{code}%
\>[0]\AgdaKeyword{variable}\<%
\\
\>[0][@{}l@{\AgdaIndent{0}}]%
\>[2]\AgdaGeneralizable{A}\AgdaSpace{}%
\AgdaGeneralizable{B}\AgdaSpace{}%
\AgdaGeneralizable{C}\AgdaSpace{}%
\AgdaSymbol{:}\AgdaSpace{}%
\AgdaDatatype{Type}\<%
\end{code}

Agda universally quantifies over any free variables named in variable
declarations

Here is a function from naturals to naturals.

\begin{code}%
\>[0]\AgdaFunction{\AgdaUnderscore{}}\AgdaSpace{}%
\AgdaSymbol{:}\AgdaSpace{}%
\AgdaDatatype{Type}\<%
\\
\>[0]\AgdaSymbol{\AgdaUnderscore{}}\AgdaSpace{}%
\AgdaSymbol{=}\AgdaSpace{}%
\AgdaInductiveConstructor{`ℕ}\AgdaSpace{}%
\AgdaOperator{\AgdaInductiveConstructor{⇒}}\AgdaSpace{}%
\AgdaInductiveConstructor{`ℕ}\<%
\end{code}

In Agda, one may use \texttt{\_} as a dummy name that is convenient for
examples.

\hypertarget{contexts}{%
\subsection{Contexts}\label{contexts}}

A context is a list of types. The type corresponding to de Bruijn index
zero appears at the right end of the list. The empty context is written
\texttt{∅}, and \texttt{\_▷\_} adds a type to the end of the list.

\begin{code}%
\>[0]\AgdaKeyword{data}\AgdaSpace{}%
\AgdaDatatype{Con}\AgdaSpace{}%
\AgdaSymbol{:}\AgdaSpace{}%
\AgdaPrimitive{Set}\AgdaSpace{}%
\AgdaKeyword{where}\<%
\\
\>[0][@{}l@{\AgdaIndent{0}}]%
\>[2]\AgdaInductiveConstructor{∅}%
\>[7]\AgdaSymbol{:}%
\>[10]\AgdaDatatype{Con}\<%
\\
\>[2]\AgdaOperator{\AgdaInductiveConstructor{\AgdaUnderscore{}▷\AgdaUnderscore{}}}%
\>[7]\AgdaSymbol{:}%
\>[10]\AgdaDatatype{Con}\AgdaSpace{}%
\AgdaSymbol{→}\AgdaSpace{}%
\AgdaDatatype{Type}\AgdaSpace{}%
\AgdaSymbol{→}\AgdaSpace{}%
\AgdaDatatype{Con}\<%
\end{code}

We let \texttt{Γ}, \texttt{Δ}, \texttt{Θ}, \texttt{Ξ} range over
environments.

\begin{code}%
\>[0]\AgdaKeyword{variable}\<%
\\
\>[0][@{}l@{\AgdaIndent{0}}]%
\>[4]\AgdaGeneralizable{Γ}\AgdaSpace{}%
\AgdaGeneralizable{Δ}\AgdaSpace{}%
\AgdaGeneralizable{Θ}\AgdaSpace{}%
\AgdaGeneralizable{Ξ}\AgdaSpace{}%
\AgdaSymbol{:}\AgdaSpace{}%
\AgdaDatatype{Con}\<%
\end{code}

Here is an environment in which de Bruijn index zero has type natural,
and de Bruijn index one is a function from naturals to naturals.

\begin{code}%
\>[0]\AgdaFunction{\AgdaUnderscore{}}\AgdaSpace{}%
\AgdaSymbol{:}\AgdaSpace{}%
\AgdaDatatype{Con}\<%
\\
\>[0]\AgdaSymbol{\AgdaUnderscore{}}\AgdaSpace{}%
\AgdaSymbol{=}\AgdaSpace{}%
\AgdaInductiveConstructor{∅}\AgdaSpace{}%
\AgdaOperator{\AgdaInductiveConstructor{▷}}\AgdaSpace{}%
\AgdaSymbol{(}\AgdaInductiveConstructor{`ℕ}\AgdaSpace{}%
\AgdaOperator{\AgdaInductiveConstructor{⇒}}\AgdaSpace{}%
\AgdaInductiveConstructor{`ℕ}\AgdaSymbol{)}\AgdaSpace{}%
\AgdaOperator{\AgdaInductiveConstructor{▷}}\AgdaSpace{}%
\AgdaInductiveConstructor{`ℕ}\<%
\end{code}

\hypertarget{terms}{%
\subsection{Terms}\label{terms}}

We write \texttt{Γ\ ⊢\ A} for the type of terms in context \texttt{Γ}
with type \texttt{A}. A term is either de Bruijn variable zero
(\texttt{●}), the weakening of a term (\texttt{M\ ↑}), a lambda
abstraction (\texttt{ƛ\ N}), an application (\texttt{L\ ·\ M}), the
number \texttt{zero}, the successor of a number \texttt{suc\ M}. (We
omit case expressions and recursion, as they add nothing to the
exposition.) Any line beginning with two dashes is a comment
What is typeset here as a rule is a line of
dashes in the source.
We take
advantage of this to make our term declarations closely resemble the
corresponding type rules.

\begin{code}%
\>[0]\AgdaKeyword{data}\AgdaSpace{}%
\AgdaOperator{\AgdaDatatype{\AgdaUnderscore{}⊢\AgdaUnderscore{}}}\AgdaSpace{}%
\AgdaSymbol{:}\AgdaSpace{}%
\AgdaDatatype{Con}\AgdaSpace{}%
\AgdaSymbol{→}\AgdaSpace{}%
\AgdaDatatype{Type}\AgdaSpace{}%
\AgdaSymbol{→}\AgdaSpace{}%
\AgdaPrimitive{Set}\AgdaSpace{}%
\AgdaKeyword{where}\<%
\\
\\[\AgdaEmptyExtraSkip]%
\>[0][@{}l@{\AgdaIndent{0}}]%
\>[2]\AgdaInductiveConstructor{●}%
\>[72I]\AgdaSymbol{:}\<%
\\
\>[72I][@{}l@{\AgdaIndent{0}}]%
\>[6]\AgdaComment{\myrule}\<%
\\
\>[6][@{}l@{\AgdaIndent{0}}]%
\>[7]\AgdaGeneralizable{Γ}\AgdaSpace{}%
\AgdaOperator{\AgdaInductiveConstructor{▷}}\AgdaSpace{}%
\AgdaGeneralizable{A}\AgdaSpace{}%
\AgdaOperator{\AgdaDatatype{⊢}}\AgdaSpace{}%
\AgdaGeneralizable{A}\<%
\\
\\[\AgdaEmptyExtraSkip]%
\>[2]\AgdaOperator{\AgdaInductiveConstructor{\AgdaUnderscore{}↑}}%
\>[77I]\AgdaSymbol{:}\<%
\\
\>[77I][@{}l@{\AgdaIndent{0}}]%
\>[6]\AgdaSymbol{(}\AgdaBound{M}\AgdaSpace{}%
\AgdaSymbol{:}\AgdaSpace{}%
\AgdaGeneralizable{Γ}\AgdaSpace{}%
\AgdaOperator{\AgdaDatatype{⊢}}\AgdaSpace{}%
\AgdaGeneralizable{B}\AgdaSymbol{)}\<%
\\
\>[2][@{}l@{\AgdaIndent{0}}]%
\>[4]\AgdaSymbol{→}%
\>[82I]\AgdaComment{\myrule}\<%
\\
\>[82I][@{}l@{\AgdaIndent{0}}]%
\>[7]\AgdaGeneralizable{Γ}\AgdaSpace{}%
\AgdaOperator{\AgdaInductiveConstructor{▷}}\AgdaSpace{}%
\AgdaGeneralizable{A}\AgdaSpace{}%
\AgdaOperator{\AgdaDatatype{⊢}}\AgdaSpace{}%
\AgdaGeneralizable{B}\<%
\\
\\[\AgdaEmptyExtraSkip]%
\>[2]\AgdaOperator{\AgdaInductiveConstructor{ƛ\AgdaUnderscore{}}}%
\>[87I]\AgdaSymbol{:}\<%
\\
\>[87I][@{}l@{\AgdaIndent{0}}]%
\>[6]\AgdaSymbol{(}\AgdaBound{N}\AgdaSpace{}%
\AgdaSymbol{:}\AgdaSpace{}%
\AgdaGeneralizable{Γ}\AgdaSpace{}%
\AgdaOperator{\AgdaInductiveConstructor{▷}}\AgdaSpace{}%
\AgdaGeneralizable{A}\AgdaSpace{}%
\AgdaOperator{\AgdaDatatype{⊢}}\AgdaSpace{}%
\AgdaGeneralizable{B}\AgdaSymbol{)}\<%
\\
\>[2][@{}l@{\AgdaIndent{0}}]%
\>[4]\AgdaSymbol{→}%
\>[94I]\AgdaComment{\myrule}\<%
\\
\>[94I][@{}l@{\AgdaIndent{0}}]%
\>[7]\AgdaGeneralizable{Γ}\AgdaSpace{}%
\AgdaOperator{\AgdaDatatype{⊢}}\AgdaSpace{}%
\AgdaGeneralizable{A}\AgdaSpace{}%
\AgdaOperator{\AgdaInductiveConstructor{⇒}}\AgdaSpace{}%
\AgdaGeneralizable{B}\<%
\\
\\[\AgdaEmptyExtraSkip]%
\>[2]\AgdaOperator{\AgdaInductiveConstructor{\AgdaUnderscore{}·\AgdaUnderscore{}}}%
\>[99I]\AgdaSymbol{:}\<%
\\
\>[.][@{}l@{}]\<[99I]%
\>[6]\AgdaSymbol{(}\AgdaBound{L}\AgdaSpace{}%
\AgdaSymbol{:}\AgdaSpace{}%
\AgdaGeneralizable{Γ}\AgdaSpace{}%
\AgdaOperator{\AgdaDatatype{⊢}}\AgdaSpace{}%
\AgdaGeneralizable{A}\AgdaSpace{}%
\AgdaOperator{\AgdaInductiveConstructor{⇒}}\AgdaSpace{}%
\AgdaGeneralizable{B}\AgdaSymbol{)}\<%
\\
\>[6]\AgdaSymbol{(}\AgdaBound{M}\AgdaSpace{}%
\AgdaSymbol{:}\AgdaSpace{}%
\AgdaGeneralizable{Γ}\AgdaSpace{}%
\AgdaOperator{\AgdaDatatype{⊢}}\AgdaSpace{}%
\AgdaGeneralizable{A}\AgdaSymbol{)}\<%
\\
\>[2][@{}l@{\AgdaIndent{0}}]%
\>[4]\AgdaSymbol{→}%
\>[110I]\AgdaComment{\myrule}\<%
\\
\>[110I][@{}l@{\AgdaIndent{0}}]%
\>[7]\AgdaGeneralizable{Γ}\AgdaSpace{}%
\AgdaOperator{\AgdaDatatype{⊢}}\AgdaSpace{}%
\AgdaGeneralizable{B}\<%
\\
\\[\AgdaEmptyExtraSkip]%
\>[2]\AgdaInductiveConstructor{zero}%
\>[113I]\AgdaSymbol{:}\<%
\\
\>[.][@{}l@{}]\<[113I]%
\>[7]\AgdaComment{\myrule}\<%
\\
\>[7][@{}l@{\AgdaIndent{0}}]%
\>[8]\AgdaGeneralizable{Γ}\AgdaSpace{}%
\AgdaOperator{\AgdaDatatype{⊢}}\AgdaSpace{}%
\AgdaInductiveConstructor{`ℕ}\<%
\\
\\[\AgdaEmptyExtraSkip]%
\>[2]\AgdaInductiveConstructor{suc}%
\>[116I]\AgdaSymbol{:}\<%
\\
\>[.][@{}l@{}]\<[116I]%
\>[6]\AgdaSymbol{(}\AgdaBound{M}\AgdaSpace{}%
\AgdaSymbol{:}\AgdaSpace{}%
\AgdaGeneralizable{Γ}\AgdaSpace{}%
\AgdaOperator{\AgdaDatatype{⊢}}\AgdaSpace{}%
\AgdaInductiveConstructor{`ℕ}\AgdaSymbol{)}\<%
\\
\>[2][@{}l@{\AgdaIndent{0}}]%
\>[4]\AgdaSymbol{→}%
\>[121I]\AgdaComment{\myrule}\<%
\\
\>[121I][@{}l@{\AgdaIndent{0}}]%
\>[7]\AgdaGeneralizable{Γ}\AgdaSpace{}%
\AgdaOperator{\AgdaDatatype{⊢}}\AgdaSpace{}%
\AgdaInductiveConstructor{`ℕ}\<%
\end{code}

We let \texttt{L}, \texttt{M}, \texttt{N}, \texttt{P}, \texttt{Q} range
over terms.

\begin{code}%
\>[0]\AgdaKeyword{variable}\<%
\\
\>[0][@{}l@{\AgdaIndent{0}}]%
\>[2]\AgdaGeneralizable{L}\AgdaSpace{}%
\AgdaGeneralizable{M}\AgdaSpace{}%
\AgdaGeneralizable{N}\AgdaSpace{}%
\AgdaGeneralizable{P}\AgdaSpace{}%
\AgdaGeneralizable{Q}\AgdaSpace{}%
\AgdaSymbol{:}\AgdaSpace{}%
\AgdaGeneralizable{Γ}\AgdaSpace{}%
\AgdaOperator{\AgdaDatatype{⊢}}\AgdaSpace{}%
\AgdaGeneralizable{A}\<%
\end{code}

Here is the increment function for naturals.

\begin{code}%
\>[0]\AgdaFunction{inc}\AgdaSpace{}%
\AgdaSymbol{:}\AgdaSpace{}%
\AgdaInductiveConstructor{∅}\AgdaSpace{}%
\AgdaOperator{\AgdaDatatype{⊢}}\AgdaSpace{}%
\AgdaInductiveConstructor{`ℕ}\AgdaSpace{}%
\AgdaOperator{\AgdaInductiveConstructor{⇒}}\AgdaSpace{}%
\AgdaInductiveConstructor{`ℕ}\<%
\\
\>[0]\AgdaFunction{inc}\AgdaSpace{}%
\AgdaSymbol{=}\AgdaSpace{}%
\AgdaOperator{\AgdaInductiveConstructor{ƛ}}\AgdaSpace{}%
\AgdaSymbol{(}\AgdaInductiveConstructor{suc}\AgdaSpace{}%
\AgdaInductiveConstructor{●}\AgdaSymbol{)}\<%
\end{code}

Here is the term for the Church numeral two.

\begin{code}%
\>[0]\AgdaFunction{two}\AgdaSpace{}%
\AgdaSymbol{:}\AgdaSpace{}%
\AgdaInductiveConstructor{∅}\AgdaSpace{}%
\AgdaOperator{\AgdaDatatype{⊢}}\AgdaSpace{}%
\AgdaSymbol{(}\AgdaGeneralizable{A}\AgdaSpace{}%
\AgdaOperator{\AgdaInductiveConstructor{⇒}}\AgdaSpace{}%
\AgdaGeneralizable{A}\AgdaSymbol{)}\AgdaSpace{}%
\AgdaOperator{\AgdaInductiveConstructor{⇒}}\AgdaSpace{}%
\AgdaGeneralizable{A}\AgdaSpace{}%
\AgdaOperator{\AgdaInductiveConstructor{⇒}}\AgdaSpace{}%
\AgdaGeneralizable{A}\<%
\\
\>[0]\AgdaFunction{two}\AgdaSpace{}%
\AgdaSymbol{=}\AgdaSpace{}%
\AgdaOperator{\AgdaInductiveConstructor{ƛ}}\AgdaSpace{}%
\AgdaSymbol{(}\AgdaOperator{\AgdaInductiveConstructor{ƛ}}\AgdaSpace{}%
\AgdaSymbol{(}\AgdaInductiveConstructor{●}\AgdaSpace{}%
\AgdaOperator{\AgdaInductiveConstructor{↑}}\AgdaSpace{}%
\AgdaOperator{\AgdaInductiveConstructor{·}}\AgdaSpace{}%
\AgdaSymbol{(}\AgdaInductiveConstructor{●}\AgdaSpace{}%
\AgdaOperator{\AgdaInductiveConstructor{↑}}\AgdaSpace{}%
\AgdaOperator{\AgdaInductiveConstructor{·}}\AgdaSpace{}%
\AgdaInductiveConstructor{●}\AgdaSymbol{)))}\<%
\end{code}

Here \texttt{●} corresponds to de Bruijn index zero, and \texttt{●\ ↑}
corresponds to de Bruijn index one.

Crucially, weakening can be applied to any term, not just a variable.
The following two open terms are equivalent.

\begin{code}%
\>[0]\AgdaFunction{M₀}\AgdaSpace{}%
\AgdaFunction{M₁}\AgdaSpace{}%
\AgdaSymbol{:}\AgdaSpace{}%
\AgdaInductiveConstructor{∅}\AgdaSpace{}%
\AgdaOperator{\AgdaInductiveConstructor{▷}}\AgdaSpace{}%
\AgdaSymbol{(}\AgdaGeneralizable{A}\AgdaSpace{}%
\AgdaOperator{\AgdaInductiveConstructor{⇒}}\AgdaSpace{}%
\AgdaGeneralizable{B}\AgdaSpace{}%
\AgdaOperator{\AgdaInductiveConstructor{⇒}}\AgdaSpace{}%
\AgdaGeneralizable{C}\AgdaSymbol{)}\AgdaSpace{}%
\AgdaOperator{\AgdaInductiveConstructor{▷}}\AgdaSpace{}%
\AgdaGeneralizable{A}\AgdaSpace{}%
\AgdaOperator{\AgdaInductiveConstructor{▷}}\AgdaSpace{}%
\AgdaGeneralizable{B}\AgdaSpace{}%
\AgdaOperator{\AgdaDatatype{⊢}}\AgdaSpace{}%
\AgdaGeneralizable{C}\<%
\\
\>[0]\AgdaFunction{M₀}\AgdaSpace{}%
\AgdaSymbol{=}\AgdaSpace{}%
\AgdaInductiveConstructor{●}\AgdaSpace{}%
\AgdaOperator{\AgdaInductiveConstructor{↑}}\AgdaSpace{}%
\AgdaOperator{\AgdaInductiveConstructor{↑}}\AgdaSpace{}%
\AgdaOperator{\AgdaInductiveConstructor{·}}\AgdaSpace{}%
\AgdaInductiveConstructor{●}\AgdaSpace{}%
\AgdaOperator{\AgdaInductiveConstructor{↑}}\AgdaSpace{}%
\AgdaOperator{\AgdaInductiveConstructor{·}}\AgdaSpace{}%
\AgdaInductiveConstructor{●}\<%
\\
\>[0]\AgdaFunction{M₁}\AgdaSpace{}%
\AgdaSymbol{=}\AgdaSpace{}%
\AgdaSymbol{(}\AgdaInductiveConstructor{●}\AgdaSpace{}%
\AgdaOperator{\AgdaInductiveConstructor{↑}}\AgdaSpace{}%
\AgdaOperator{\AgdaInductiveConstructor{·}}\AgdaSpace{}%
\AgdaInductiveConstructor{●}\AgdaSymbol{)}\AgdaSpace{}%
\AgdaOperator{\AgdaInductiveConstructor{↑}}\AgdaSpace{}%
\AgdaOperator{\AgdaInductiveConstructor{·}}\AgdaSpace{}%
\AgdaInductiveConstructor{●}\<%
\end{code}

Here \texttt{●\ ↑\ ↑} corresponds to de Bruijn index two.

\hypertarget{substitutions}{%
\subsection{Substitutions}\label{substitutions}}

We write \texttt{Γ\ ⊨\ Δ} for the type of a substitution that replaces
variables in environment \texttt{Δ} by terms in environment \texttt{Γ}.
A substitution is either the identity substitution \texttt{id}, the
weakening of a substitution \texttt{σ\ ↑}, or the cons of a substitution
and a term \texttt{σ\ ▷\ P}. We take advantage of overloading to permit
weakening on both terms \texttt{M\ ↑} and substitutions \texttt{σ\ ↑},
and to permit cons on both environments \texttt{Γ\ ▷\ A} and
substitutions \texttt{σ\ ▷\ P}.

\begin{code}%
\>[0]\AgdaKeyword{data}\AgdaSpace{}%
\AgdaOperator{\AgdaDatatype{\AgdaUnderscore{}⊨\AgdaUnderscore{}}}\AgdaSpace{}%
\AgdaSymbol{:}\AgdaSpace{}%
\AgdaDatatype{Con}\AgdaSpace{}%
\AgdaSymbol{→}\AgdaSpace{}%
\AgdaDatatype{Con}\AgdaSpace{}%
\AgdaSymbol{→}\AgdaSpace{}%
\AgdaPrimitive{Set}\AgdaSpace{}%
\AgdaKeyword{where}\<%
\\
\\[\AgdaEmptyExtraSkip]%
\>[0][@{}l@{\AgdaIndent{0}}]%
\>[2]\AgdaInductiveConstructor{id}%
\>[202I]\AgdaSymbol{:}\<%
\\
\>[202I][@{}l@{\AgdaIndent{0}}]%
\>[6]\AgdaComment{\myrule}\<%
\\
\>[6][@{}l@{\AgdaIndent{0}}]%
\>[7]\AgdaGeneralizable{Δ}\AgdaSpace{}%
\AgdaOperator{\AgdaDatatype{⊨}}\AgdaSpace{}%
\AgdaGeneralizable{Δ}\<%
\\
\\[\AgdaEmptyExtraSkip]%
\>[2]\AgdaOperator{\AgdaInductiveConstructor{\AgdaUnderscore{}↑}}%
\>[205I]\AgdaSymbol{:}\<%
\\
\>[205I][@{}l@{\AgdaIndent{0}}]%
\>[6]\AgdaSymbol{(}\AgdaBound{σ}\AgdaSpace{}%
\AgdaSymbol{:}\AgdaSpace{}%
\AgdaGeneralizable{Γ}\AgdaSpace{}%
\AgdaOperator{\AgdaDatatype{⊨}}\AgdaSpace{}%
\AgdaGeneralizable{Δ}\AgdaSymbol{)}\<%
\\
\>[2][@{}l@{\AgdaIndent{0}}]%
\>[4]\AgdaSymbol{→}%
\>[210I]\AgdaComment{\myrule}\<%
\\
\>[210I][@{}l@{\AgdaIndent{0}}]%
\>[7]\AgdaGeneralizable{Γ}\AgdaSpace{}%
\AgdaOperator{\AgdaInductiveConstructor{▷}}\AgdaSpace{}%
\AgdaGeneralizable{A}\AgdaSpace{}%
\AgdaOperator{\AgdaDatatype{⊨}}\AgdaSpace{}%
\AgdaGeneralizable{Δ}\<%
\\
\\[\AgdaEmptyExtraSkip]%
\>[2]\AgdaOperator{\AgdaInductiveConstructor{\AgdaUnderscore{}▷\AgdaUnderscore{}}}%
\>[215I]\AgdaSymbol{:}\<%
\\
\>[.][@{}l@{}]\<[215I]%
\>[6]\AgdaSymbol{(}\AgdaBound{σ}\AgdaSpace{}%
\AgdaSymbol{:}\AgdaSpace{}%
\AgdaGeneralizable{Γ}\AgdaSpace{}%
\AgdaOperator{\AgdaDatatype{⊨}}\AgdaSpace{}%
\AgdaGeneralizable{Δ}\AgdaSymbol{)}\<%
\\
\>[6]\AgdaSymbol{(}\AgdaBound{M}\AgdaSpace{}%
\AgdaSymbol{:}\AgdaSpace{}%
\AgdaGeneralizable{Γ}\AgdaSpace{}%
\AgdaOperator{\AgdaDatatype{⊢}}\AgdaSpace{}%
\AgdaGeneralizable{A}\AgdaSymbol{)}\<%
\\
\>[2][@{}l@{\AgdaIndent{0}}]%
\>[4]\AgdaSymbol{→}%
\>[224I]\AgdaComment{\myrule}\<%
\\
\>[224I][@{}l@{\AgdaIndent{0}}]%
\>[7]\AgdaGeneralizable{Γ}\AgdaSpace{}%
\AgdaOperator{\AgdaDatatype{⊨}}\AgdaSpace{}%
\AgdaGeneralizable{Δ}\AgdaSpace{}%
\AgdaOperator{\AgdaInductiveConstructor{▷}}\AgdaSpace{}%
\AgdaGeneralizable{A}\<%
\end{code}

We let \texttt{σ}, \texttt{τ}, \texttt{υ} range over substitutions.

\begin{code}%
\>[0]\AgdaKeyword{variable}\<%
\\
\>[0][@{}l@{\AgdaIndent{0}}]%
\>[2]\AgdaGeneralizable{σ}\AgdaSpace{}%
\AgdaGeneralizable{τ}\AgdaSpace{}%
\AgdaGeneralizable{υ}\AgdaSpace{}%
\AgdaSymbol{:}\AgdaSpace{}%
\AgdaGeneralizable{Γ}\AgdaSpace{}%
\AgdaOperator{\AgdaDatatype{⊨}}\AgdaSpace{}%
\AgdaGeneralizable{Δ}\<%
\end{code}

We can think of a substitution \texttt{Γ\ ⊨\ Δ} built with repeated uses
of cons as a list of terms in context \texttt{Γ}, with one term for each
type in \texttt{Δ}. For example, here is a substitution that replaces de
Bruijn index zero by the number zero, and de Bruijn index one by the
increment function on naturals. Each term in the substitution is a
closed term, as indicated by the source of the substitution being the
empty environment. Here \texttt{id} has type \texttt{∅\ ⊨\ ∅}, as there
are no other free variables.

\begin{code}%
\>[0]\AgdaFunction{\AgdaUnderscore{}}\AgdaSpace{}%
\AgdaSymbol{:}\AgdaSpace{}%
\AgdaInductiveConstructor{∅}\AgdaSpace{}%
\AgdaOperator{\AgdaDatatype{⊨}}\AgdaSpace{}%
\AgdaInductiveConstructor{∅}\AgdaSpace{}%
\AgdaOperator{\AgdaInductiveConstructor{▷}}\AgdaSpace{}%
\AgdaSymbol{(}\AgdaInductiveConstructor{`ℕ}\AgdaSpace{}%
\AgdaOperator{\AgdaInductiveConstructor{⇒}}\AgdaSpace{}%
\AgdaInductiveConstructor{`ℕ}\AgdaSymbol{)}\AgdaSpace{}%
\AgdaOperator{\AgdaInductiveConstructor{▷}}\AgdaSpace{}%
\AgdaInductiveConstructor{`ℕ}\<%
\\
\>[0]\AgdaSymbol{\AgdaUnderscore{}}\AgdaSpace{}%
\AgdaSymbol{=}\AgdaSpace{}%
\AgdaInductiveConstructor{id}\AgdaSpace{}%
\AgdaOperator{\AgdaInductiveConstructor{▷}}\AgdaSpace{}%
\AgdaFunction{inc}\AgdaSpace{}%
\AgdaOperator{\AgdaInductiveConstructor{▷}}\AgdaSpace{}%
\AgdaInductiveConstructor{zero}\<%
\end{code}

Here is a substitution that replaces de Bruijn index one by the
weakening of the increment function, and leaves de Bruijn index zero
unchanged.

\begin{code}%
\>[0]\AgdaFunction{\AgdaUnderscore{}}\AgdaSpace{}%
\AgdaSymbol{:}\AgdaSpace{}%
\AgdaInductiveConstructor{∅}\AgdaSpace{}%
\AgdaOperator{\AgdaInductiveConstructor{▷}}\AgdaSpace{}%
\AgdaInductiveConstructor{`ℕ}\AgdaSpace{}%
\AgdaOperator{\AgdaDatatype{⊨}}\AgdaSpace{}%
\AgdaInductiveConstructor{∅}\AgdaSpace{}%
\AgdaOperator{\AgdaInductiveConstructor{▷}}\AgdaSpace{}%
\AgdaInductiveConstructor{`ℕ}\AgdaSpace{}%
\AgdaOperator{\AgdaInductiveConstructor{⇒}}\AgdaSpace{}%
\AgdaInductiveConstructor{`ℕ}\AgdaSpace{}%
\AgdaOperator{\AgdaInductiveConstructor{▷}}\AgdaSpace{}%
\AgdaInductiveConstructor{`ℕ}\<%
\\
\>[0]\AgdaSymbol{\AgdaUnderscore{}}\AgdaSpace{}%
\AgdaSymbol{=}\AgdaSpace{}%
\AgdaSymbol{(}\AgdaInductiveConstructor{id}\AgdaSpace{}%
\AgdaOperator{\AgdaInductiveConstructor{▷}}\AgdaSpace{}%
\AgdaFunction{inc}\AgdaSymbol{)}\AgdaSpace{}%
\AgdaOperator{\AgdaInductiveConstructor{↑}}\AgdaSpace{}%
\AgdaOperator{\AgdaInductiveConstructor{▷}}\AgdaSpace{}%
\AgdaInductiveConstructor{●}\<%
\end{code}

If we omitted the weakening, the substitution would not be well typed.
This is an advantage of the intrinsic approach: correctly maintaining de
Bruijn indices is notoriously tricky, but with intrinsic typing getting
a de Bruijn index wrong usually leads to a type error.

Finally, here is a substitution that flips two variables, making the
outermost innermost and vice versa.

\begin{code}%
\>[0]\AgdaFunction{\AgdaUnderscore{}}\AgdaSpace{}%
\AgdaSymbol{:}\AgdaSpace{}%
\AgdaInductiveConstructor{∅}\AgdaSpace{}%
\AgdaOperator{\AgdaInductiveConstructor{▷}}\AgdaSpace{}%
\AgdaGeneralizable{A}\AgdaSpace{}%
\AgdaOperator{\AgdaInductiveConstructor{▷}}\AgdaSpace{}%
\AgdaGeneralizable{B}\AgdaSpace{}%
\AgdaOperator{\AgdaDatatype{⊨}}\AgdaSpace{}%
\AgdaInductiveConstructor{∅}\AgdaSpace{}%
\AgdaOperator{\AgdaInductiveConstructor{▷}}\AgdaSpace{}%
\AgdaGeneralizable{B}\AgdaSpace{}%
\AgdaOperator{\AgdaInductiveConstructor{▷}}\AgdaSpace{}%
\AgdaGeneralizable{A}\<%
\\
\>[0]\AgdaSymbol{\AgdaUnderscore{}}\AgdaSpace{}%
\AgdaSymbol{=}\AgdaSpace{}%
\AgdaInductiveConstructor{id}\AgdaSpace{}%
\AgdaOperator{\AgdaInductiveConstructor{↑}}\AgdaSpace{}%
\AgdaOperator{\AgdaInductiveConstructor{↑}}\AgdaSpace{}%
\AgdaOperator{\AgdaInductiveConstructor{▷}}\AgdaSpace{}%
\AgdaInductiveConstructor{●}\AgdaSpace{}%
\AgdaOperator{\AgdaInductiveConstructor{▷}}\AgdaSpace{}%
\AgdaInductiveConstructor{●}\AgdaSpace{}%
\AgdaOperator{\AgdaInductiveConstructor{↑}}\<%
\end{code}

Here \texttt{id\ ↑\ ↑\ :\ ∅\ ▷\ A\ ▷\ B\ ⊨\ ∅}, as required by the type
rules for cons. Again, the types keep us straight. If we accidentally
add or drop an \texttt{↑} the term will become ill-typed.

Often, we will need to know that a substitution is a cons, but will want
to refer to the whole substitution. For this purpose, it is convenient
to use a pattern declaration to declare \texttt{△} as a shorthand for
\texttt{\_\ ▷\ \_}.

\begin{code}%
\>[0]\AgdaKeyword{pattern}\AgdaSpace{}%
\AgdaInductiveConstructor{△}%
\>[11]\AgdaSymbol{=}%
\>[14]\AgdaSymbol{\AgdaUnderscore{}}\AgdaSpace{}%
\AgdaOperator{\AgdaInductiveConstructor{▷}}\AgdaSpace{}%
\AgdaSymbol{\AgdaUnderscore{}}\<%
\end{code}

We can then pattern match against \texttt{(σ\ @\ △)}, which binds
\texttt{σ} to a substitution, but only if it is in the form of a cons.

\hypertarget{instantiation}{%
\subsection{Instantiation}\label{instantiation}}

We write \texttt{M\ {[}\ σ\ {]}} to instantiate term \texttt{M} with
substitution \texttt{σ}. Instantiation is contravariant: substitution
\texttt{σ\ :\ Γ\ ⊨\ Δ} takes a term \texttt{M\ :\ Δ\ ⊢\ A} into a term
\texttt{M\ {[}\ σ\ {]}\ :\ Γ\ ⊢\ A}.

\begin{code}%
\>[0]\AgdaOperator{\AgdaFunction{\AgdaUnderscore{}[\AgdaUnderscore{}]}}\AgdaSpace{}%
\AgdaSymbol{:}\<%
\\
\>[0][@{}l@{\AgdaIndent{0}}]%
\>[4]\AgdaSymbol{(}\AgdaBound{M}\AgdaSpace{}%
\AgdaSymbol{:}\AgdaSpace{}%
\AgdaGeneralizable{Δ}\AgdaSpace{}%
\AgdaOperator{\AgdaDatatype{⊢}}\AgdaSpace{}%
\AgdaGeneralizable{A}\AgdaSymbol{)}\<%
\\
\>[4]\AgdaSymbol{(}\AgdaBound{σ}\AgdaSpace{}%
\AgdaSymbol{:}\AgdaSpace{}%
\AgdaGeneralizable{Γ}\AgdaSpace{}%
\AgdaOperator{\AgdaDatatype{⊨}}\AgdaSpace{}%
\AgdaGeneralizable{Δ}\AgdaSymbol{)}\<%
\\
\>[0][@{}l@{\AgdaIndent{0}}]%
\>[2]\AgdaSymbol{→}%
\>[303I]\AgdaComment{\myrule}\<%
\\
\>[303I][@{}l@{\AgdaIndent{0}}]%
\>[5]\AgdaGeneralizable{Γ}\AgdaSpace{}%
\AgdaOperator{\AgdaDatatype{⊢}}\AgdaSpace{}%
\AgdaGeneralizable{A}\<%
\\
\>[0]\AgdaBound{M}%
\>[14]\AgdaOperator{\AgdaFunction{[}}\AgdaSpace{}%
\AgdaInductiveConstructor{id}\AgdaSpace{}%
\AgdaOperator{\AgdaFunction{]}}%
\>[25]\AgdaSymbol{=}%
\>[28]\AgdaBound{M}%
\>[47]\AgdaComment{--\ (1)}\<%
\\
\>[0]\AgdaBound{M}%
\>[14]\AgdaOperator{\AgdaFunction{[}}\AgdaSpace{}%
\AgdaBound{σ}\AgdaSpace{}%
\AgdaOperator{\AgdaInductiveConstructor{↑}}\AgdaSpace{}%
\AgdaOperator{\AgdaFunction{]}}%
\>[25]\AgdaSymbol{=}%
\>[28]\AgdaSymbol{(}\AgdaBound{M}\AgdaSpace{}%
\AgdaOperator{\AgdaFunction{[}}\AgdaSpace{}%
\AgdaBound{σ}\AgdaSpace{}%
\AgdaOperator{\AgdaFunction{]}}\AgdaSymbol{)}\AgdaSpace{}%
\AgdaOperator{\AgdaInductiveConstructor{↑}}%
\>[47]\AgdaComment{--\ (2)}\<%
\\
\>[0]\AgdaInductiveConstructor{●}%
\>[14]\AgdaOperator{\AgdaFunction{[}}\AgdaSpace{}%
\AgdaBound{σ}\AgdaSpace{}%
\AgdaOperator{\AgdaInductiveConstructor{▷}}\AgdaSpace{}%
\AgdaBound{P}\AgdaSpace{}%
\AgdaOperator{\AgdaFunction{]}}%
\>[25]\AgdaSymbol{=}%
\>[28]\AgdaBound{P}%
\>[47]\AgdaComment{--\ (3)}\<%
\\
\>[0]\AgdaSymbol{(}\AgdaBound{M}\AgdaSpace{}%
\AgdaOperator{\AgdaInductiveConstructor{↑}}\AgdaSymbol{)}%
\>[14]\AgdaOperator{\AgdaFunction{[}}\AgdaSpace{}%
\AgdaBound{σ}\AgdaSpace{}%
\AgdaOperator{\AgdaInductiveConstructor{▷}}\AgdaSpace{}%
\AgdaBound{P}\AgdaSpace{}%
\AgdaOperator{\AgdaFunction{]}}%
\>[25]\AgdaSymbol{=}%
\>[28]\AgdaBound{M}\AgdaSpace{}%
\AgdaOperator{\AgdaFunction{[}}\AgdaSpace{}%
\AgdaBound{σ}\AgdaSpace{}%
\AgdaOperator{\AgdaFunction{]}}%
\>[47]\AgdaComment{--\ (4)}\<%
\\
\>[0]\AgdaSymbol{(}\AgdaOperator{\AgdaInductiveConstructor{ƛ}}\AgdaSpace{}%
\AgdaBound{N}\AgdaSymbol{)}%
\>[14]\AgdaOperator{\AgdaFunction{[}}\AgdaSpace{}%
\AgdaBound{σ}\AgdaSpace{}%
\AgdaSymbol{@}\AgdaSpace{}%
\AgdaInductiveConstructor{△}\AgdaSpace{}%
\AgdaOperator{\AgdaFunction{]}}%
\>[25]\AgdaSymbol{=}%
\>[28]\AgdaOperator{\AgdaInductiveConstructor{ƛ}}\AgdaSpace{}%
\AgdaSymbol{(}\AgdaBound{N}\AgdaSpace{}%
\AgdaOperator{\AgdaFunction{[}}\AgdaSpace{}%
\AgdaBound{σ}\AgdaSpace{}%
\AgdaOperator{\AgdaInductiveConstructor{↑}}\AgdaSpace{}%
\AgdaOperator{\AgdaInductiveConstructor{▷}}\AgdaSpace{}%
\AgdaInductiveConstructor{●}\AgdaSpace{}%
\AgdaOperator{\AgdaFunction{]}}\AgdaSymbol{)}%
\>[47]\AgdaComment{--\ (5)}\<%
\\
\>[0]\AgdaSymbol{(}\AgdaBound{L}\AgdaSpace{}%
\AgdaOperator{\AgdaInductiveConstructor{·}}\AgdaSpace{}%
\AgdaBound{M}\AgdaSymbol{)}%
\>[14]\AgdaOperator{\AgdaFunction{[}}\AgdaSpace{}%
\AgdaBound{σ}\AgdaSpace{}%
\AgdaSymbol{@}\AgdaSpace{}%
\AgdaInductiveConstructor{△}\AgdaSpace{}%
\AgdaOperator{\AgdaFunction{]}}%
\>[25]\AgdaSymbol{=}%
\>[28]\AgdaBound{L}\AgdaSpace{}%
\AgdaOperator{\AgdaFunction{[}}\AgdaSpace{}%
\AgdaBound{σ}\AgdaSpace{}%
\AgdaOperator{\AgdaFunction{]}}\AgdaSpace{}%
\AgdaOperator{\AgdaInductiveConstructor{·}}\AgdaSpace{}%
\AgdaBound{M}\AgdaSpace{}%
\AgdaOperator{\AgdaFunction{[}}\AgdaSpace{}%
\AgdaBound{σ}\AgdaSpace{}%
\AgdaOperator{\AgdaFunction{]}}%
\>[47]\AgdaComment{--\ (6)}\<%
\\
\>[0]\AgdaInductiveConstructor{zero}%
\>[14]\AgdaOperator{\AgdaFunction{[}}\AgdaSpace{}%
\AgdaBound{σ}\AgdaSpace{}%
\AgdaSymbol{@}\AgdaSpace{}%
\AgdaInductiveConstructor{△}\AgdaSpace{}%
\AgdaOperator{\AgdaFunction{]}}%
\>[25]\AgdaSymbol{=}%
\>[28]\AgdaInductiveConstructor{zero}%
\>[47]\AgdaComment{--\ (7)}\<%
\\
\>[0]\AgdaSymbol{(}\AgdaInductiveConstructor{suc}\AgdaSpace{}%
\AgdaBound{M}\AgdaSymbol{)}%
\>[14]\AgdaOperator{\AgdaFunction{[}}\AgdaSpace{}%
\AgdaBound{σ}\AgdaSpace{}%
\AgdaSymbol{@}\AgdaSpace{}%
\AgdaInductiveConstructor{△}\AgdaSpace{}%
\AgdaOperator{\AgdaFunction{]}}%
\>[25]\AgdaSymbol{=}%
\>[28]\AgdaInductiveConstructor{suc}\AgdaSpace{}%
\AgdaSymbol{(}\AgdaBound{M}\AgdaSpace{}%
\AgdaOperator{\AgdaFunction{[}}\AgdaSpace{}%
\AgdaBound{σ}\AgdaSpace{}%
\AgdaOperator{\AgdaFunction{]}}\AgdaSymbol{)}%
\>[47]\AgdaComment{--\ (8)}\<%
\end{code}

Instantiation is defined by case analysis. It is crucial that we first
perform case analysis on the substitution, since if it is identity or
weakening we can compute the result easily without considering the
structure of a term---that is the whole point of representing identity
and weakening explicitly! Only if the substitution is a cons do we need
to perform a case analysis on the term.

Thus, we first do case analysis on the substitution. (1) For identity
the definition is obvious. (2) Weakening of a substitution becomes
weakening of a term---this is why we defined explicit weakening in the
first place. If the substitution is a cons, we then do a case analysis
on the term. (3) If the term is de Bruijn variable zero we take the term
from the cons. (4) If the term is a weakening we drop the term from the
cons and apply recursively. Otherwise, we push the substitution into the
term. (5) If the term is a lambda abstraction, the substitution
\texttt{σ} becomes \texttt{σ\ ↑\ ▷\ ●} as it is pushed under the lambda.
Bound variable zero is left unchanged, while variables one and up now
map to the terms bound by \texttt{σ}, weakened to account for the newly
intervening lambda binder. Our notation neatly encapsulates all
renumbering of de Bruijn indexes, which is usually considered tricky.
(6--8) Otherwise, the substitution is unchanged and applied recursively
to each part. This completes the definition of instantiation.

Beta reduction is written as follows.

\begin{verbatim}
(ƛ N) · M  —→  N [ id ▷ M ]
\end{verbatim}

Substution \texttt{id\ ▷\ M} maps variable zero to \texttt{M}, and maps
variable one to zero, two to one, and so on, as required since the
surrounding lambda binder has been removed. Once again, our notation
neatly encodes the tricky renumbering of de Bruijn indexes.

Consider an application of the Church numeral two to the increment
function.

\begin{verbatim}
two · inc
\end{verbatim}

Beta reduction yields an application of a substitution, which we compute
as follows. Each equality is labelled with its justifying line from the
definition.

\begin{code}%
\>[0]\AgdaFunction{\AgdaUnderscore{}}\AgdaSpace{}%
\AgdaSymbol{:}\AgdaSpace{}%
\AgdaSymbol{(}\AgdaOperator{\AgdaInductiveConstructor{ƛ}}\AgdaSpace{}%
\AgdaSymbol{(}\AgdaInductiveConstructor{●}\AgdaSpace{}%
\AgdaOperator{\AgdaInductiveConstructor{↑}}\AgdaSpace{}%
\AgdaOperator{\AgdaInductiveConstructor{·}}\AgdaSpace{}%
\AgdaSymbol{(}\AgdaInductiveConstructor{●}\AgdaSpace{}%
\AgdaOperator{\AgdaInductiveConstructor{↑}}\AgdaSpace{}%
\AgdaOperator{\AgdaInductiveConstructor{·}}\AgdaSpace{}%
\AgdaInductiveConstructor{●}\AgdaSymbol{)))}\AgdaSpace{}%
\AgdaOperator{\AgdaFunction{[}}\AgdaSpace{}%
\AgdaInductiveConstructor{id}\AgdaSpace{}%
\AgdaOperator{\AgdaInductiveConstructor{▷}}\AgdaSpace{}%
\AgdaFunction{inc}\AgdaSpace{}%
\AgdaOperator{\AgdaFunction{]}}%
\>[40]\AgdaOperator{\AgdaDatatype{≡}}%
\>[43]\AgdaOperator{\AgdaInductiveConstructor{ƛ}}\AgdaSpace{}%
\AgdaSymbol{(}\AgdaFunction{inc}\AgdaSpace{}%
\AgdaOperator{\AgdaInductiveConstructor{↑}}\AgdaSpace{}%
\AgdaOperator{\AgdaInductiveConstructor{·}}\AgdaSpace{}%
\AgdaSymbol{(}\AgdaFunction{inc}\AgdaSpace{}%
\AgdaOperator{\AgdaInductiveConstructor{↑}}\AgdaSpace{}%
\AgdaOperator{\AgdaInductiveConstructor{·}}\AgdaSpace{}%
\AgdaInductiveConstructor{●}\AgdaSymbol{))}\<%
\\
\>[0]\AgdaSymbol{\AgdaUnderscore{}}%
\>[387I]\AgdaSymbol{=}\<%
\\
\>[387I][@{}l@{\AgdaIndent{0}}]%
\>[4]\AgdaOperator{\AgdaFunction{begin}}\<%
\\
\>[4][@{}l@{\AgdaIndent{0}}]%
\>[6]\AgdaSymbol{(}\AgdaOperator{\AgdaInductiveConstructor{ƛ}}\AgdaSpace{}%
\AgdaSymbol{(}\AgdaInductiveConstructor{●}\AgdaSpace{}%
\AgdaOperator{\AgdaInductiveConstructor{↑}}\AgdaSpace{}%
\AgdaOperator{\AgdaInductiveConstructor{·}}\AgdaSpace{}%
\AgdaSymbol{(}\AgdaInductiveConstructor{●}\AgdaSpace{}%
\AgdaOperator{\AgdaInductiveConstructor{↑}}\AgdaSpace{}%
\AgdaOperator{\AgdaInductiveConstructor{·}}\AgdaSpace{}%
\AgdaInductiveConstructor{●}\AgdaSymbol{)))}\AgdaSpace{}%
\AgdaOperator{\AgdaFunction{[}}\AgdaSpace{}%
\AgdaInductiveConstructor{id}\AgdaSpace{}%
\AgdaOperator{\AgdaInductiveConstructor{▷}}\AgdaSpace{}%
\AgdaFunction{inc}\AgdaSpace{}%
\AgdaOperator{\AgdaFunction{]}}\<%
\\
\>[4]\AgdaFunction{≡⟨⟩}%
\>[16]\AgdaComment{--\ (5)}\<%
\\
\>[4][@{}l@{\AgdaIndent{0}}]%
\>[6]\AgdaOperator{\AgdaInductiveConstructor{ƛ}}\AgdaSpace{}%
\AgdaSymbol{((}\AgdaInductiveConstructor{●}\AgdaSpace{}%
\AgdaOperator{\AgdaInductiveConstructor{↑}}\AgdaSpace{}%
\AgdaOperator{\AgdaInductiveConstructor{·}}\AgdaSpace{}%
\AgdaSymbol{(}\AgdaInductiveConstructor{●}\AgdaSpace{}%
\AgdaOperator{\AgdaInductiveConstructor{↑}}\AgdaSpace{}%
\AgdaOperator{\AgdaInductiveConstructor{·}}\AgdaSpace{}%
\AgdaInductiveConstructor{●}\AgdaSymbol{))}\AgdaSpace{}%
\AgdaOperator{\AgdaFunction{[}}\AgdaSpace{}%
\AgdaSymbol{(}\AgdaInductiveConstructor{id}\AgdaSpace{}%
\AgdaOperator{\AgdaInductiveConstructor{▷}}\AgdaSpace{}%
\AgdaFunction{inc}\AgdaSymbol{)}\AgdaSpace{}%
\AgdaOperator{\AgdaInductiveConstructor{↑}}\AgdaSpace{}%
\AgdaOperator{\AgdaInductiveConstructor{▷}}\AgdaSpace{}%
\AgdaInductiveConstructor{●}\AgdaSpace{}%
\AgdaOperator{\AgdaFunction{]}}\AgdaSymbol{)}\<%
\\
\>[4]\AgdaFunction{≡⟨⟩}%
\>[16]\AgdaComment{--\ (6)}\<%
\\
\>[4][@{}l@{\AgdaIndent{0}}]%
\>[6]\AgdaOperator{\AgdaInductiveConstructor{ƛ}}\AgdaSpace{}%
\AgdaSymbol{(((}\AgdaInductiveConstructor{●}\AgdaSpace{}%
\AgdaOperator{\AgdaInductiveConstructor{↑}}\AgdaSymbol{)}\AgdaSpace{}%
\AgdaOperator{\AgdaFunction{[}}\AgdaSpace{}%
\AgdaSymbol{(}\AgdaInductiveConstructor{id}\AgdaSpace{}%
\AgdaOperator{\AgdaInductiveConstructor{▷}}\AgdaSpace{}%
\AgdaFunction{inc}\AgdaSymbol{)}\AgdaSpace{}%
\AgdaOperator{\AgdaInductiveConstructor{↑}}\AgdaSpace{}%
\AgdaOperator{\AgdaInductiveConstructor{▷}}\AgdaSpace{}%
\AgdaInductiveConstructor{●}\AgdaSpace{}%
\AgdaOperator{\AgdaFunction{]}}\AgdaSymbol{)}\AgdaSpace{}%
\AgdaOperator{\AgdaInductiveConstructor{·}}\AgdaSpace{}%
\AgdaSymbol{((}\AgdaInductiveConstructor{●}\AgdaSpace{}%
\AgdaOperator{\AgdaInductiveConstructor{↑}}\AgdaSpace{}%
\AgdaOperator{\AgdaInductiveConstructor{·}}\AgdaSpace{}%
\AgdaInductiveConstructor{●}\AgdaSymbol{)}\AgdaSpace{}%
\AgdaOperator{\AgdaFunction{[}}\AgdaSpace{}%
\AgdaSymbol{(}\AgdaInductiveConstructor{id}\AgdaSpace{}%
\AgdaOperator{\AgdaInductiveConstructor{▷}}\AgdaSpace{}%
\AgdaFunction{inc}\AgdaSymbol{)}\AgdaSpace{}%
\AgdaOperator{\AgdaInductiveConstructor{↑}}\AgdaSpace{}%
\AgdaOperator{\AgdaInductiveConstructor{▷}}\AgdaSpace{}%
\AgdaInductiveConstructor{●}\AgdaSpace{}%
\AgdaOperator{\AgdaFunction{]}}\AgdaSymbol{))}\<%
\\
\>[4]\AgdaFunction{≡⟨⟩}%
\>[16]\AgdaComment{--\ (4)}\<%
\\
\>[4][@{}l@{\AgdaIndent{0}}]%
\>[6]\AgdaOperator{\AgdaInductiveConstructor{ƛ}}\AgdaSpace{}%
\AgdaSymbol{((}\AgdaInductiveConstructor{●}\AgdaSpace{}%
\AgdaOperator{\AgdaFunction{[}}\AgdaSpace{}%
\AgdaSymbol{(}\AgdaInductiveConstructor{id}\AgdaSpace{}%
\AgdaOperator{\AgdaInductiveConstructor{▷}}\AgdaSpace{}%
\AgdaFunction{inc}\AgdaSymbol{)}\AgdaSpace{}%
\AgdaOperator{\AgdaInductiveConstructor{↑}}\AgdaSpace{}%
\AgdaOperator{\AgdaFunction{]}}\AgdaSymbol{)}\AgdaSpace{}%
\AgdaOperator{\AgdaInductiveConstructor{·}}\AgdaSpace{}%
\AgdaSymbol{((}\AgdaInductiveConstructor{●}\AgdaSpace{}%
\AgdaOperator{\AgdaInductiveConstructor{↑}}\AgdaSpace{}%
\AgdaOperator{\AgdaInductiveConstructor{·}}\AgdaSpace{}%
\AgdaInductiveConstructor{●}\AgdaSymbol{)}\AgdaSpace{}%
\AgdaOperator{\AgdaFunction{[}}\AgdaSpace{}%
\AgdaSymbol{(}\AgdaInductiveConstructor{id}\AgdaSpace{}%
\AgdaOperator{\AgdaInductiveConstructor{▷}}\AgdaSpace{}%
\AgdaFunction{inc}\AgdaSymbol{)}\AgdaSpace{}%
\AgdaOperator{\AgdaInductiveConstructor{↑}}\AgdaSpace{}%
\AgdaOperator{\AgdaInductiveConstructor{▷}}\AgdaSpace{}%
\AgdaInductiveConstructor{●}\AgdaSpace{}%
\AgdaOperator{\AgdaFunction{]}}\AgdaSymbol{))}\<%
\\
\>[4]\AgdaFunction{≡⟨⟩}%
\>[16]\AgdaComment{--\ (2)}\<%
\\
\>[4][@{}l@{\AgdaIndent{0}}]%
\>[6]\AgdaOperator{\AgdaInductiveConstructor{ƛ}}\AgdaSpace{}%
\AgdaSymbol{((}\AgdaInductiveConstructor{●}\AgdaSpace{}%
\AgdaOperator{\AgdaFunction{[}}\AgdaSpace{}%
\AgdaInductiveConstructor{id}\AgdaSpace{}%
\AgdaOperator{\AgdaInductiveConstructor{▷}}\AgdaSpace{}%
\AgdaFunction{inc}\AgdaSpace{}%
\AgdaOperator{\AgdaFunction{]}}\AgdaSymbol{)}\AgdaSpace{}%
\AgdaOperator{\AgdaInductiveConstructor{↑}}\AgdaSpace{}%
\AgdaOperator{\AgdaInductiveConstructor{·}}\AgdaSpace{}%
\AgdaSymbol{((}\AgdaInductiveConstructor{●}\AgdaSpace{}%
\AgdaOperator{\AgdaInductiveConstructor{↑}}\AgdaSpace{}%
\AgdaOperator{\AgdaInductiveConstructor{·}}\AgdaSpace{}%
\AgdaInductiveConstructor{●}\AgdaSymbol{)}\AgdaSpace{}%
\AgdaOperator{\AgdaFunction{[}}\AgdaSpace{}%
\AgdaSymbol{(}\AgdaInductiveConstructor{id}\AgdaSpace{}%
\AgdaOperator{\AgdaInductiveConstructor{▷}}\AgdaSpace{}%
\AgdaFunction{inc}\AgdaSymbol{)}\AgdaSpace{}%
\AgdaOperator{\AgdaInductiveConstructor{↑}}\AgdaSpace{}%
\AgdaOperator{\AgdaInductiveConstructor{▷}}\AgdaSpace{}%
\AgdaInductiveConstructor{●}\AgdaSpace{}%
\AgdaOperator{\AgdaFunction{]}}\AgdaSymbol{))}\<%
\\
\>[4]\AgdaFunction{≡⟨⟩}%
\>[16]\AgdaComment{--\ (3)}\<%
\\
\>[4][@{}l@{\AgdaIndent{0}}]%
\>[6]\AgdaOperator{\AgdaInductiveConstructor{ƛ}}\AgdaSpace{}%
\AgdaSymbol{(}\AgdaFunction{inc}\AgdaSpace{}%
\AgdaOperator{\AgdaInductiveConstructor{↑}}\AgdaSpace{}%
\AgdaOperator{\AgdaInductiveConstructor{·}}\AgdaSpace{}%
\AgdaSymbol{((}\AgdaInductiveConstructor{●}\AgdaSpace{}%
\AgdaOperator{\AgdaInductiveConstructor{↑}}\AgdaSpace{}%
\AgdaOperator{\AgdaInductiveConstructor{·}}\AgdaSpace{}%
\AgdaInductiveConstructor{●}\AgdaSymbol{)}\AgdaSpace{}%
\AgdaOperator{\AgdaFunction{[}}\AgdaSpace{}%
\AgdaSymbol{(}\AgdaInductiveConstructor{id}\AgdaSpace{}%
\AgdaOperator{\AgdaInductiveConstructor{▷}}\AgdaSpace{}%
\AgdaFunction{inc}\AgdaSymbol{)}\AgdaSpace{}%
\AgdaOperator{\AgdaInductiveConstructor{↑}}\AgdaSpace{}%
\AgdaOperator{\AgdaInductiveConstructor{▷}}\AgdaSpace{}%
\AgdaInductiveConstructor{●}\AgdaSpace{}%
\AgdaOperator{\AgdaFunction{]}}\AgdaSymbol{))}\<%
\\
\>[4]\AgdaFunction{≡⟨⟩}%
\>[16]\AgdaComment{--\ \ ...}\<%
\\
\>[4][@{}l@{\AgdaIndent{0}}]%
\>[6]\AgdaOperator{\AgdaInductiveConstructor{ƛ}}\AgdaSpace{}%
\AgdaSymbol{(}\AgdaFunction{inc}\AgdaSpace{}%
\AgdaOperator{\AgdaInductiveConstructor{↑}}\AgdaSpace{}%
\AgdaOperator{\AgdaInductiveConstructor{·}}\AgdaSpace{}%
\AgdaSymbol{(}\AgdaFunction{inc}\AgdaSpace{}%
\AgdaOperator{\AgdaInductiveConstructor{↑}}\AgdaSpace{}%
\AgdaOperator{\AgdaInductiveConstructor{·}}\AgdaSpace{}%
\AgdaInductiveConstructor{●}\AgdaSymbol{))}\<%
\\
\>[4]\AgdaOperator{\AgdaFunction{∎}}\<%
\end{code}

Since \texttt{inc} is a closed term, it must be weakened to appear
underneath a lambda, and this is accomplished by the rule that converts
\texttt{σ} to \texttt{σ\ ↑\ ▷\ ●} when pushing a substitution under a
lambda abstraction. The end of the computation adds nothing new, so some
details are omitted. Indeed all the detail can be omitted, as Agda can
confirm the result simply by normalising both sides.

\begin{code}%
\>[0]\AgdaFunction{\AgdaUnderscore{}}\AgdaSpace{}%
\AgdaSymbol{:}\AgdaSpace{}%
\AgdaSymbol{(}\AgdaOperator{\AgdaInductiveConstructor{ƛ}}\AgdaSpace{}%
\AgdaSymbol{(}\AgdaInductiveConstructor{●}\AgdaSpace{}%
\AgdaOperator{\AgdaInductiveConstructor{↑}}\AgdaSpace{}%
\AgdaOperator{\AgdaInductiveConstructor{·}}\AgdaSpace{}%
\AgdaSymbol{(}\AgdaInductiveConstructor{●}\AgdaSpace{}%
\AgdaOperator{\AgdaInductiveConstructor{↑}}\AgdaSpace{}%
\AgdaOperator{\AgdaInductiveConstructor{·}}\AgdaSpace{}%
\AgdaInductiveConstructor{●}\AgdaSymbol{)))}\AgdaSpace{}%
\AgdaOperator{\AgdaFunction{[}}\AgdaSpace{}%
\AgdaInductiveConstructor{id}\AgdaSpace{}%
\AgdaOperator{\AgdaInductiveConstructor{▷}}\AgdaSpace{}%
\AgdaFunction{inc}\AgdaSpace{}%
\AgdaOperator{\AgdaFunction{]}}%
\>[40]\AgdaOperator{\AgdaDatatype{≡}}%
\>[43]\AgdaOperator{\AgdaInductiveConstructor{ƛ}}\AgdaSpace{}%
\AgdaSymbol{(}\AgdaFunction{inc}\AgdaSpace{}%
\AgdaOperator{\AgdaInductiveConstructor{↑}}\AgdaSpace{}%
\AgdaOperator{\AgdaInductiveConstructor{·}}\AgdaSpace{}%
\AgdaSymbol{(}\AgdaFunction{inc}\AgdaSpace{}%
\AgdaOperator{\AgdaInductiveConstructor{↑}}\AgdaSpace{}%
\AgdaOperator{\AgdaInductiveConstructor{·}}\AgdaSpace{}%
\AgdaInductiveConstructor{●}\AgdaSymbol{))}\<%
\\
\>[0]\AgdaSymbol{\AgdaUnderscore{}}\AgdaSpace{}%
\AgdaSymbol{=}\AgdaSpace{}%
\AgdaInductiveConstructor{refl}\<%
\end{code}

\hypertarget{composition}{%
\subsection{Composition}\label{composition}}

We write \texttt{σ\ ⨟\ τ} for composition of substitutions. If
\texttt{σ\ :\ Θ\ ⊨\ Δ} and \texttt{τ\ :\ Γ\ ⊨\ Θ} then
\texttt{(σ\ ⨟\ τ)\ :\ Γ\ ⇛\ Δ}.

\begin{code}%
\>[0]\AgdaOperator{\AgdaFunction{\AgdaUnderscore{}⨾\AgdaUnderscore{}}}%
\>[523I]\AgdaSymbol{:}\<%
\\
\>[.][@{}l@{}]\<[523I]%
\>[4]\AgdaSymbol{(}\AgdaBound{σ}\AgdaSpace{}%
\AgdaSymbol{:}\AgdaSpace{}%
\AgdaGeneralizable{Θ}\AgdaSpace{}%
\AgdaOperator{\AgdaDatatype{⊨}}\AgdaSpace{}%
\AgdaGeneralizable{Δ}\AgdaSymbol{)}\<%
\\
\>[4]\AgdaSymbol{(}\AgdaBound{τ}\AgdaSpace{}%
\AgdaSymbol{:}\AgdaSpace{}%
\AgdaGeneralizable{Γ}\AgdaSpace{}%
\AgdaOperator{\AgdaDatatype{⊨}}\AgdaSpace{}%
\AgdaGeneralizable{Θ}\AgdaSymbol{)}\<%
\\
\>[0][@{}l@{\AgdaIndent{0}}]%
\>[2]\AgdaSymbol{→}%
\>[532I]\AgdaComment{\myrule}\<%
\\
\>[532I][@{}l@{\AgdaIndent{0}}]%
\>[5]\AgdaGeneralizable{Γ}\AgdaSpace{}%
\AgdaOperator{\AgdaDatatype{⊨}}\AgdaSpace{}%
\AgdaGeneralizable{Δ}\<%
\\
\>[0]\AgdaBound{σ}%
\>[9]\AgdaOperator{\AgdaFunction{⨾}}%
\>[12]\AgdaInductiveConstructor{id}%
\>[21]\AgdaSymbol{=}%
\>[24]\AgdaBound{σ}%
\>[45]\AgdaComment{--\ (1)}\<%
\\
\>[0]\AgdaBound{σ}%
\>[9]\AgdaOperator{\AgdaFunction{⨾}}%
\>[12]\AgdaSymbol{(}\AgdaBound{τ}\AgdaSpace{}%
\AgdaOperator{\AgdaInductiveConstructor{↑}}\AgdaSymbol{)}%
\>[21]\AgdaSymbol{=}%
\>[24]\AgdaSymbol{(}\AgdaBound{σ}\AgdaSpace{}%
\AgdaOperator{\AgdaFunction{⨾}}\AgdaSpace{}%
\AgdaBound{τ}\AgdaSymbol{)}\AgdaSpace{}%
\AgdaOperator{\AgdaInductiveConstructor{↑}}%
\>[45]\AgdaComment{--\ (2)}\<%
\\
\>[0]\AgdaInductiveConstructor{id}%
\>[9]\AgdaOperator{\AgdaFunction{⨾}}%
\>[12]\AgdaSymbol{(}\AgdaBound{τ}\AgdaSpace{}%
\AgdaSymbol{@}\AgdaSpace{}%
\AgdaInductiveConstructor{△}\AgdaSymbol{)}%
\>[21]\AgdaSymbol{=}%
\>[24]\AgdaBound{τ}%
\>[45]\AgdaComment{--\ (3)}\<%
\\
\>[0]\AgdaSymbol{(}\AgdaBound{σ}\AgdaSpace{}%
\AgdaOperator{\AgdaInductiveConstructor{↑}}\AgdaSymbol{)}%
\>[9]\AgdaOperator{\AgdaFunction{⨾}}%
\>[12]\AgdaSymbol{(}\AgdaBound{τ}\AgdaSpace{}%
\AgdaOperator{\AgdaInductiveConstructor{▷}}\AgdaSpace{}%
\AgdaBound{Q}\AgdaSymbol{)}%
\>[21]\AgdaSymbol{=}%
\>[24]\AgdaBound{σ}\AgdaSpace{}%
\AgdaOperator{\AgdaFunction{⨾}}\AgdaSpace{}%
\AgdaBound{τ}%
\>[45]\AgdaComment{--\ (4)}\<%
\\
\>[0]\AgdaSymbol{(}\AgdaBound{σ}\AgdaSpace{}%
\AgdaOperator{\AgdaInductiveConstructor{▷}}\AgdaSpace{}%
\AgdaBound{P}\AgdaSymbol{)}%
\>[9]\AgdaOperator{\AgdaFunction{⨾}}%
\>[12]\AgdaSymbol{(}\AgdaBound{τ}\AgdaSpace{}%
\AgdaSymbol{@}\AgdaSpace{}%
\AgdaInductiveConstructor{△}\AgdaSymbol{)}%
\>[21]\AgdaSymbol{=}%
\>[24]\AgdaSymbol{(}\AgdaBound{σ}\AgdaSpace{}%
\AgdaOperator{\AgdaFunction{⨾}}\AgdaSpace{}%
\AgdaBound{τ}\AgdaSymbol{)}\AgdaSpace{}%
\AgdaOperator{\AgdaInductiveConstructor{▷}}\AgdaSpace{}%
\AgdaSymbol{(}\AgdaBound{P}\AgdaSpace{}%
\AgdaOperator{\AgdaFunction{[}}\AgdaSpace{}%
\AgdaBound{τ}\AgdaSpace{}%
\AgdaOperator{\AgdaFunction{]}}\AgdaSymbol{)}%
\>[45]\AgdaComment{--\ (5)}\<%
\end{code}

The case analysis for instantiation goes right-to-left: first we analyse
the substitution, and only if it is a cons do we analyse the term.
Hence, composition is also defined right-to-left: first we analyse the
right substitution \texttt{τ}, and only if it is a cons do we analyse
the left substitution \texttt{σ}. Each of the equations should be
familiar by now.

\hypertarget{composition-and-instantiation}{%
\subsection{Composition and
instantiation}\label{composition-and-instantiation}}

A key result relates composition and instantiation.

\begin{code}%
\>[0]\AgdaFunction{[][]}\AgdaSpace{}%
\AgdaSymbol{:}\<%
\\
\>[0][@{}l@{\AgdaIndent{0}}]%
\>[4]\AgdaSymbol{(}\AgdaBound{M}\AgdaSpace{}%
\AgdaSymbol{:}\AgdaSpace{}%
\AgdaGeneralizable{Δ}\AgdaSpace{}%
\AgdaOperator{\AgdaDatatype{⊢}}\AgdaSpace{}%
\AgdaGeneralizable{A}\AgdaSymbol{)}\<%
\\
\>[4]\AgdaSymbol{(}\AgdaBound{σ}\AgdaSpace{}%
\AgdaSymbol{:}\AgdaSpace{}%
\AgdaGeneralizable{Θ}\AgdaSpace{}%
\AgdaOperator{\AgdaDatatype{⊨}}\AgdaSpace{}%
\AgdaGeneralizable{Δ}\AgdaSymbol{)}\<%
\\
\>[4]\AgdaSymbol{(}\AgdaBound{τ}\AgdaSpace{}%
\AgdaSymbol{:}\AgdaSpace{}%
\AgdaGeneralizable{Γ}\AgdaSpace{}%
\AgdaOperator{\AgdaDatatype{⊨}}\AgdaSpace{}%
\AgdaGeneralizable{Θ}\AgdaSymbol{)}\<%
\\
\>[0][@{}l@{\AgdaIndent{0}}]%
\>[2]\AgdaSymbol{→}%
\>[570I]\AgdaComment{\myrule}\<%
\\
\>[570I][@{}l@{\AgdaIndent{0}}]%
\>[5]\AgdaBound{M}\AgdaSpace{}%
\AgdaOperator{\AgdaFunction{[}}\AgdaSpace{}%
\AgdaBound{σ}\AgdaSpace{}%
\AgdaOperator{\AgdaFunction{]}}\AgdaSpace{}%
\AgdaOperator{\AgdaFunction{[}}\AgdaSpace{}%
\AgdaBound{τ}\AgdaSpace{}%
\AgdaOperator{\AgdaFunction{]}}\AgdaSpace{}%
\AgdaOperator{\AgdaDatatype{≡}}\AgdaSpace{}%
\AgdaBound{M}\AgdaSpace{}%
\AgdaOperator{\AgdaFunction{[}}\AgdaSpace{}%
\AgdaBound{σ}\AgdaSpace{}%
\AgdaOperator{\AgdaFunction{⨾}}\AgdaSpace{}%
\AgdaBound{τ}\AgdaSpace{}%
\AgdaOperator{\AgdaFunction{]}}\<%
\end{code}

The only tricky step is to recognise that case analysis of the arguments
must proceed right-to-left, to match the definitions of application and
composition. Hence, first we perform case analysis on \texttt{τ}, and
only if \texttt{τ} is a cons do we perform case analysis on \texttt{σ},
and only if both are conses do we perform case analysis on \texttt{M}.
Below we use the operator \texttt{cong}, short for congruence. If
\texttt{eq} proves \texttt{x\ ≡\ y}, then \texttt{cong\ f\ eq} proves
\texttt{f\ x\ ≡\ f\ y}. Similarly for \texttt{cong₂}.

\begin{code}%
\>[0]\AgdaFunction{[][]}%
\>[6]\AgdaBound{M}%
\>[20]\AgdaBound{σ}%
\>[29]\AgdaInductiveConstructor{id}%
\>[38]\AgdaSymbol{=}%
\>[41]\AgdaInductiveConstructor{refl}%
\>[79]\AgdaComment{--\ \ (1)}\<%
\\
\>[0]\AgdaFunction{[][]}%
\>[6]\AgdaBound{M}%
\>[20]\AgdaBound{σ}%
\>[29]\AgdaSymbol{(}\AgdaBound{τ}\AgdaSpace{}%
\AgdaOperator{\AgdaInductiveConstructor{↑}}\AgdaSymbol{)}%
\>[38]\AgdaSymbol{=}%
\>[41]\AgdaFunction{cong}\AgdaSpace{}%
\AgdaOperator{\AgdaInductiveConstructor{\AgdaUnderscore{}↑}}\AgdaSpace{}%
\AgdaSymbol{(}\AgdaFunction{[][]}\AgdaSpace{}%
\AgdaBound{M}\AgdaSpace{}%
\AgdaBound{σ}\AgdaSpace{}%
\AgdaBound{τ}\AgdaSymbol{)}%
\>[79]\AgdaComment{--\ \ (2)}\<%
\\
\>[0]\AgdaFunction{[][]}%
\>[6]\AgdaBound{M}%
\>[20]\AgdaInductiveConstructor{id}%
\>[29]\AgdaSymbol{(}\AgdaBound{τ}\AgdaSpace{}%
\AgdaSymbol{@}\AgdaSpace{}%
\AgdaInductiveConstructor{△}\AgdaSymbol{)}%
\>[38]\AgdaSymbol{=}%
\>[41]\AgdaInductiveConstructor{refl}%
\>[79]\AgdaComment{--\ \ (3)}\<%
\\
\>[0]\AgdaFunction{[][]}%
\>[6]\AgdaBound{M}%
\>[20]\AgdaSymbol{(}\AgdaBound{σ}\AgdaSpace{}%
\AgdaOperator{\AgdaInductiveConstructor{↑}}\AgdaSymbol{)}%
\>[29]\AgdaSymbol{(}\AgdaBound{τ}\AgdaSpace{}%
\AgdaOperator{\AgdaInductiveConstructor{▷}}\AgdaSpace{}%
\AgdaBound{Q}\AgdaSymbol{)}%
\>[38]\AgdaSymbol{=}%
\>[41]\AgdaFunction{[][]}\AgdaSpace{}%
\AgdaBound{M}\AgdaSpace{}%
\AgdaBound{σ}\AgdaSpace{}%
\AgdaBound{τ}%
\>[79]\AgdaComment{--\ \ (4)}\<%
\\
\>[0]\AgdaFunction{[][]}%
\>[6]\AgdaInductiveConstructor{●}%
\>[20]\AgdaSymbol{(}\AgdaBound{σ}\AgdaSpace{}%
\AgdaOperator{\AgdaInductiveConstructor{▷}}\AgdaSpace{}%
\AgdaBound{P}\AgdaSymbol{)}%
\>[29]\AgdaSymbol{(}\AgdaBound{τ}\AgdaSpace{}%
\AgdaSymbol{@}\AgdaSpace{}%
\AgdaInductiveConstructor{△}\AgdaSymbol{)}%
\>[38]\AgdaSymbol{=}%
\>[41]\AgdaInductiveConstructor{refl}%
\>[79]\AgdaComment{--\ \ (5)}\<%
\\
\>[0]\AgdaFunction{[][]}%
\>[6]\AgdaSymbol{(}\AgdaBound{M}\AgdaSpace{}%
\AgdaOperator{\AgdaInductiveConstructor{↑}}\AgdaSymbol{)}%
\>[20]\AgdaSymbol{(}\AgdaBound{σ}\AgdaSpace{}%
\AgdaOperator{\AgdaInductiveConstructor{▷}}\AgdaSpace{}%
\AgdaBound{P}\AgdaSymbol{)}%
\>[29]\AgdaSymbol{(}\AgdaBound{τ}\AgdaSpace{}%
\AgdaSymbol{@}\AgdaSpace{}%
\AgdaInductiveConstructor{△}\AgdaSymbol{)}%
\>[38]\AgdaSymbol{=}%
\>[41]\AgdaFunction{[][]}\AgdaSpace{}%
\AgdaBound{M}\AgdaSpace{}%
\AgdaBound{σ}\AgdaSpace{}%
\AgdaBound{τ}%
\>[79]\AgdaComment{--\ \ (6)}\<%
\\
\>[0]\AgdaFunction{[][]}%
\>[6]\AgdaSymbol{(}\AgdaOperator{\AgdaInductiveConstructor{ƛ}}\AgdaSpace{}%
\AgdaBound{N}\AgdaSymbol{)}%
\>[20]\AgdaSymbol{(}\AgdaBound{σ}\AgdaSpace{}%
\AgdaSymbol{@}\AgdaSpace{}%
\AgdaInductiveConstructor{△}\AgdaSymbol{)}%
\>[29]\AgdaSymbol{(}\AgdaBound{τ}\AgdaSpace{}%
\AgdaSymbol{@}\AgdaSpace{}%
\AgdaInductiveConstructor{△}\AgdaSymbol{)}%
\>[38]\AgdaSymbol{=}%
\>[41]\AgdaFunction{cong}\AgdaSpace{}%
\AgdaOperator{\AgdaInductiveConstructor{ƛ\AgdaUnderscore{}}}\AgdaSpace{}%
\AgdaSymbol{(}\AgdaFunction{[][]}\AgdaSpace{}%
\AgdaBound{N}\AgdaSpace{}%
\AgdaSymbol{(}\AgdaBound{σ}\AgdaSpace{}%
\AgdaOperator{\AgdaInductiveConstructor{↑}}\AgdaSpace{}%
\AgdaOperator{\AgdaInductiveConstructor{▷}}\AgdaSpace{}%
\AgdaInductiveConstructor{●}\AgdaSymbol{)}\AgdaSpace{}%
\AgdaSymbol{(}\AgdaBound{τ}\AgdaSpace{}%
\AgdaOperator{\AgdaInductiveConstructor{↑}}\AgdaSpace{}%
\AgdaOperator{\AgdaInductiveConstructor{▷}}\AgdaSpace{}%
\AgdaInductiveConstructor{●}\AgdaSymbol{))}%
\>[79]\AgdaComment{--\ \ (7)}\<%
\\
\>[0]\AgdaFunction{[][]}%
\>[6]\AgdaSymbol{(}\AgdaBound{L}\AgdaSpace{}%
\AgdaOperator{\AgdaInductiveConstructor{·}}\AgdaSpace{}%
\AgdaBound{M}\AgdaSymbol{)}%
\>[20]\AgdaSymbol{(}\AgdaBound{σ}\AgdaSpace{}%
\AgdaSymbol{@}\AgdaSpace{}%
\AgdaInductiveConstructor{△}\AgdaSymbol{)}%
\>[29]\AgdaSymbol{(}\AgdaBound{τ}\AgdaSpace{}%
\AgdaSymbol{@}\AgdaSpace{}%
\AgdaInductiveConstructor{△}\AgdaSymbol{)}%
\>[38]\AgdaSymbol{=}%
\>[41]\AgdaFunction{cong₂}\AgdaSpace{}%
\AgdaOperator{\AgdaInductiveConstructor{\AgdaUnderscore{}·\AgdaUnderscore{}}}\AgdaSpace{}%
\AgdaSymbol{(}\AgdaFunction{[][]}\AgdaSpace{}%
\AgdaBound{L}\AgdaSpace{}%
\AgdaBound{σ}\AgdaSpace{}%
\AgdaBound{τ}\AgdaSymbol{)}\AgdaSpace{}%
\AgdaSymbol{(}\AgdaFunction{[][]}\AgdaSpace{}%
\AgdaBound{M}\AgdaSpace{}%
\AgdaBound{σ}\AgdaSpace{}%
\AgdaBound{τ}\AgdaSymbol{)}%
\>[79]\AgdaComment{--\ \ (8)}\<%
\\
\>[0]\AgdaFunction{[][]}%
\>[6]\AgdaInductiveConstructor{zero}%
\>[20]\AgdaSymbol{(}\AgdaBound{σ}\AgdaSpace{}%
\AgdaSymbol{@}\AgdaSpace{}%
\AgdaInductiveConstructor{△}\AgdaSymbol{)}%
\>[29]\AgdaSymbol{(}\AgdaBound{τ}\AgdaSpace{}%
\AgdaSymbol{@}\AgdaSpace{}%
\AgdaInductiveConstructor{△}\AgdaSymbol{)}%
\>[38]\AgdaSymbol{=}%
\>[41]\AgdaInductiveConstructor{refl}%
\>[79]\AgdaComment{--\ \ (9)}\<%
\\
\>[0]\AgdaFunction{[][]}%
\>[6]\AgdaSymbol{(}\AgdaInductiveConstructor{suc}\AgdaSpace{}%
\AgdaBound{M}\AgdaSymbol{)}%
\>[20]\AgdaSymbol{(}\AgdaBound{σ}\AgdaSpace{}%
\AgdaSymbol{@}\AgdaSpace{}%
\AgdaInductiveConstructor{△}\AgdaSymbol{)}%
\>[29]\AgdaSymbol{(}\AgdaBound{τ}\AgdaSpace{}%
\AgdaSymbol{@}\AgdaSpace{}%
\AgdaInductiveConstructor{△}\AgdaSymbol{)}%
\>[38]\AgdaSymbol{=}%
\>[41]\AgdaFunction{cong}\AgdaSpace{}%
\AgdaInductiveConstructor{suc}\AgdaSpace{}%
\AgdaSymbol{(}\AgdaFunction{[][]}\AgdaSpace{}%
\AgdaBound{M}\AgdaSpace{}%
\AgdaBound{σ}\AgdaSpace{}%
\AgdaBound{τ}\AgdaSymbol{)}%
\>[79]\AgdaComment{--\ (10)}\<%
\end{code}

For instance, for (2) the two sides simplify to

\begin{verbatim}
M [ σ ] [ τ ] ↑  ≡  M [ σ ⨟ τ ] ↑
\end{verbatim}

and the equation follows by an application of the induction hypothesis.
Similarly, for (8) the two sides simplify to

\begin{verbatim}
(L [ σ ] [ τ ]) · (M [ σ ] [ τ ]) ≡ (L [ σ ⨟ τ ]) · (M [ σ ⨟ τ ])
\end{verbatim}

and the equation follows by two applications of the induction
hypothesis. For (7), we push substitutions underneath a lambda, giving

\begin{verbatim}
ƛ (N [ σ ↑ ▷ ● ] [ τ ↑ ▷ ● ]) ≡ N [ (σ ↑ ▷ ●) ⨾ (τ ↑ ▷ ●) ] ≡ N [ (σ ⨾ τ) ↑ ▷ ● ]
\end{verbatim}

where the first equivalence follows by the induction hypothesis, but
applied to the substitutions \texttt{σ\ ↑\ ▷\ ●} and
\texttt{τ\ ↑\ ▷\ ●}. It is fine under structural induction for the
substitutions to get larger so long as the term is getting smaller. The
second equivalence follows by straightforward computation.

Having proved the lemma, we can now instruct Agda to apply it as a
left-to-right rewrite whenever possible when simplifying a term. This
will play a key role in reducing equations of interest to a triviality.

\begin{code}%
\>[0]\AgdaSymbol{\{-\#}\AgdaSpace{}%
\AgdaKeyword{REWRITE}\AgdaSpace{}%
\AgdaFunction{[][]}\AgdaSpace{}%
\AgdaSymbol{\#-\}}\<%
\end{code}

\hypertarget{composition-has-a-left-identity.}{%
\subsection{Composition has a left
identity.}\label{composition-has-a-left-identity.}}

Composition has \texttt{id} as a right identity by definition. It is
easy to show that it is also a left identity.

\begin{code}%
\>[0]\AgdaFunction{left-id}\AgdaSpace{}%
\AgdaSymbol{:}\<%
\\
\>[0][@{}l@{\AgdaIndent{0}}]%
\>[4]\AgdaSymbol{(}\AgdaBound{τ}\AgdaSpace{}%
\AgdaSymbol{:}\AgdaSpace{}%
\AgdaGeneralizable{Γ}\AgdaSpace{}%
\AgdaOperator{\AgdaDatatype{⊨}}\AgdaSpace{}%
\AgdaGeneralizable{Δ}\AgdaSymbol{)}\<%
\\
\>[0][@{}l@{\AgdaIndent{0}}]%
\>[2]\AgdaSymbol{→}%
\>[663I]\AgdaComment{\myrule}\<%
\\
\>[663I][@{}l@{\AgdaIndent{0}}]%
\>[5]\AgdaInductiveConstructor{id}\AgdaSpace{}%
\AgdaOperator{\AgdaFunction{⨾}}\AgdaSpace{}%
\AgdaBound{τ}\AgdaSpace{}%
\AgdaOperator{\AgdaDatatype{≡}}\AgdaSpace{}%
\AgdaBound{τ}\<%
\\
\>[0]\AgdaFunction{left-id}\AgdaSpace{}%
\AgdaInductiveConstructor{id}%
\>[17]\AgdaSymbol{=}%
\>[20]\AgdaInductiveConstructor{refl}%
\>[41]\AgdaComment{--\ (1)}\<%
\\
\>[0]\AgdaFunction{left-id}\AgdaSpace{}%
\AgdaSymbol{(}\AgdaBound{τ}\AgdaSpace{}%
\AgdaOperator{\AgdaInductiveConstructor{↑}}\AgdaSymbol{)}%
\>[17]\AgdaSymbol{=}%
\>[20]\AgdaFunction{cong}\AgdaSpace{}%
\AgdaOperator{\AgdaInductiveConstructor{\AgdaUnderscore{}↑}}\AgdaSpace{}%
\AgdaSymbol{(}\AgdaFunction{left-id}\AgdaSpace{}%
\AgdaBound{τ}\AgdaSymbol{)}%
\>[41]\AgdaComment{--\ (2)}\<%
\\
\>[0]\AgdaFunction{left-id}\AgdaSpace{}%
\AgdaSymbol{(}\AgdaBound{τ}\AgdaSpace{}%
\AgdaOperator{\AgdaInductiveConstructor{▷}}\AgdaSpace{}%
\AgdaBound{Q}\AgdaSymbol{)}%
\>[17]\AgdaSymbol{=}%
\>[20]\AgdaInductiveConstructor{refl}%
\>[41]\AgdaComment{--\ (3)}\<%
\end{code}

Obviously, the case analysis is on \texttt{τ}. (1, 3): Both sides
simplify to the same term. (2): The two sides simplify to

\begin{verbatim}
(id ⨾ τ) ↑  ≡  τ ↑
\end{verbatim}

and the result follows by the induction hypothesis.

We direct Agda to apply left identity as a rewrite.

\begin{code}%
\>[0]\AgdaSymbol{\{-\#}\AgdaSpace{}%
\AgdaKeyword{REWRITE}\AgdaSpace{}%
\AgdaFunction{left-id}\AgdaSpace{}%
\AgdaSymbol{\#-\}}\<%
\end{code}

\hypertarget{composition-is-associative}{%
\subsection{Composition is
associative}\label{composition-is-associative}}

We can also show that composition is associative.

\begin{code}%
\>[0]\AgdaFunction{assoc}\AgdaSpace{}%
\AgdaSymbol{:}\<%
\\
\>[0][@{}l@{\AgdaIndent{0}}]%
\>[4]\AgdaSymbol{(}\AgdaBound{σ}\AgdaSpace{}%
\AgdaSymbol{:}\AgdaSpace{}%
\AgdaGeneralizable{Θ}\AgdaSpace{}%
\AgdaOperator{\AgdaDatatype{⊨}}\AgdaSpace{}%
\AgdaGeneralizable{Δ}\AgdaSymbol{)}\<%
\\
\>[4]\AgdaSymbol{(}\AgdaBound{τ}\AgdaSpace{}%
\AgdaSymbol{:}\AgdaSpace{}%
\AgdaGeneralizable{Ξ}\AgdaSpace{}%
\AgdaOperator{\AgdaDatatype{⊨}}\AgdaSpace{}%
\AgdaGeneralizable{Θ}\AgdaSymbol{)}\<%
\\
\>[4]\AgdaSymbol{(}\AgdaBound{υ}\AgdaSpace{}%
\AgdaSymbol{:}\AgdaSpace{}%
\AgdaGeneralizable{Γ}\AgdaSpace{}%
\AgdaOperator{\AgdaDatatype{⊨}}\AgdaSpace{}%
\AgdaGeneralizable{Ξ}\AgdaSymbol{)}\<%
\\
\>[0][@{}l@{\AgdaIndent{0}}]%
\>[2]\AgdaSymbol{→}%
\>[693I]\AgdaComment{\myrule}\<%
\\
\>[693I][@{}l@{\AgdaIndent{0}}]%
\>[5]\AgdaSymbol{(}\AgdaBound{σ}\AgdaSpace{}%
\AgdaOperator{\AgdaFunction{⨾}}\AgdaSpace{}%
\AgdaBound{τ}\AgdaSymbol{)}\AgdaSpace{}%
\AgdaOperator{\AgdaFunction{⨾}}\AgdaSpace{}%
\AgdaBound{υ}\AgdaSpace{}%
\AgdaOperator{\AgdaDatatype{≡}}\AgdaSpace{}%
\AgdaBound{σ}\AgdaSpace{}%
\AgdaOperator{\AgdaFunction{⨾}}\AgdaSpace{}%
\AgdaSymbol{(}\AgdaBound{τ}\AgdaSpace{}%
\AgdaOperator{\AgdaFunction{⨾}}\AgdaSpace{}%
\AgdaBound{υ}\AgdaSymbol{)}\<%
\\
\>[0]\AgdaFunction{assoc}%
\>[7]\AgdaBound{σ}%
\>[16]\AgdaBound{τ}%
\>[25]\AgdaInductiveConstructor{id}%
\>[34]\AgdaSymbol{=}%
\>[37]\AgdaInductiveConstructor{refl}%
\>[79]\AgdaComment{--\ (1)}\<%
\\
\>[0]\AgdaFunction{assoc}%
\>[7]\AgdaBound{σ}%
\>[16]\AgdaBound{τ}%
\>[25]\AgdaSymbol{(}\AgdaBound{υ}\AgdaSpace{}%
\AgdaOperator{\AgdaInductiveConstructor{↑}}\AgdaSymbol{)}%
\>[34]\AgdaSymbol{=}%
\>[37]\AgdaFunction{cong}\AgdaSpace{}%
\AgdaOperator{\AgdaInductiveConstructor{\AgdaUnderscore{}↑}}\AgdaSpace{}%
\AgdaSymbol{(}\AgdaFunction{assoc}\AgdaSpace{}%
\AgdaBound{σ}\AgdaSpace{}%
\AgdaBound{τ}\AgdaSpace{}%
\AgdaBound{υ}\AgdaSymbol{)}%
\>[79]\AgdaComment{--\ (2)}\<%
\\
\>[0]\AgdaFunction{assoc}%
\>[7]\AgdaBound{σ}%
\>[16]\AgdaInductiveConstructor{id}%
\>[25]\AgdaSymbol{(}\AgdaBound{υ}\AgdaSpace{}%
\AgdaOperator{\AgdaInductiveConstructor{▷}}\AgdaSpace{}%
\AgdaBound{R}\AgdaSymbol{)}%
\>[34]\AgdaSymbol{=}%
\>[37]\AgdaInductiveConstructor{refl}%
\>[79]\AgdaComment{--\ (3)}\<%
\\
\>[0]\AgdaFunction{assoc}%
\>[7]\AgdaBound{σ}%
\>[16]\AgdaSymbol{(}\AgdaBound{τ}\AgdaSpace{}%
\AgdaOperator{\AgdaInductiveConstructor{↑}}\AgdaSymbol{)}%
\>[25]\AgdaSymbol{(}\AgdaBound{υ}\AgdaSpace{}%
\AgdaOperator{\AgdaInductiveConstructor{▷}}\AgdaSpace{}%
\AgdaBound{R}\AgdaSymbol{)}%
\>[34]\AgdaSymbol{=}%
\>[37]\AgdaFunction{assoc}\AgdaSpace{}%
\AgdaBound{σ}\AgdaSpace{}%
\AgdaBound{τ}\AgdaSpace{}%
\AgdaBound{υ}%
\>[79]\AgdaComment{--\ (4)}\<%
\\
\>[0]\AgdaFunction{assoc}%
\>[7]\AgdaInductiveConstructor{id}%
\>[16]\AgdaSymbol{(}\AgdaBound{τ}\AgdaSpace{}%
\AgdaOperator{\AgdaInductiveConstructor{▷}}\AgdaSpace{}%
\AgdaBound{Q}\AgdaSymbol{)}%
\>[25]\AgdaSymbol{(}\AgdaBound{υ}\AgdaSpace{}%
\AgdaOperator{\AgdaInductiveConstructor{▷}}\AgdaSpace{}%
\AgdaBound{R}\AgdaSymbol{)}%
\>[34]\AgdaSymbol{=}%
\>[37]\AgdaInductiveConstructor{refl}%
\>[79]\AgdaComment{--\ (5)}\<%
\\
\>[0]\AgdaFunction{assoc}%
\>[7]\AgdaSymbol{(}\AgdaBound{σ}\AgdaSpace{}%
\AgdaOperator{\AgdaInductiveConstructor{↑}}\AgdaSymbol{)}%
\>[16]\AgdaSymbol{(}\AgdaBound{τ}\AgdaSpace{}%
\AgdaOperator{\AgdaInductiveConstructor{▷}}\AgdaSpace{}%
\AgdaBound{Q}\AgdaSymbol{)}%
\>[25]\AgdaSymbol{(}\AgdaBound{υ}\AgdaSpace{}%
\AgdaOperator{\AgdaInductiveConstructor{▷}}\AgdaSpace{}%
\AgdaBound{R}\AgdaSymbol{)}%
\>[34]\AgdaSymbol{=}%
\>[37]\AgdaFunction{assoc}\AgdaSpace{}%
\AgdaBound{σ}\AgdaSpace{}%
\AgdaBound{τ}\AgdaSpace{}%
\AgdaSymbol{(}\AgdaBound{υ}\AgdaSpace{}%
\AgdaOperator{\AgdaInductiveConstructor{▷}}\AgdaSpace{}%
\AgdaBound{R}\AgdaSymbol{)}%
\>[79]\AgdaComment{--\ (6)}\<%
\\
\>[0]\AgdaFunction{assoc}%
\>[7]\AgdaSymbol{(}\AgdaBound{σ}\AgdaSpace{}%
\AgdaOperator{\AgdaInductiveConstructor{▷}}\AgdaSpace{}%
\AgdaBound{P}\AgdaSymbol{)}%
\>[16]\AgdaSymbol{(}\AgdaBound{τ}\AgdaSpace{}%
\AgdaOperator{\AgdaInductiveConstructor{▷}}\AgdaSpace{}%
\AgdaBound{Q}\AgdaSymbol{)}%
\>[25]\AgdaSymbol{(}\AgdaBound{υ}\AgdaSpace{}%
\AgdaOperator{\AgdaInductiveConstructor{▷}}\AgdaSpace{}%
\AgdaBound{R}\AgdaSymbol{)}%
\>[34]\AgdaSymbol{=}%
\>[37]\AgdaFunction{cong₂}\AgdaSpace{}%
\AgdaOperator{\AgdaInductiveConstructor{\AgdaUnderscore{}▷\AgdaUnderscore{}}}\AgdaSpace{}%
\AgdaSymbol{(}\AgdaFunction{assoc}\AgdaSpace{}%
\AgdaBound{σ}\AgdaSpace{}%
\AgdaSymbol{(}\AgdaBound{τ}\AgdaSpace{}%
\AgdaOperator{\AgdaInductiveConstructor{▷}}\AgdaSpace{}%
\AgdaBound{Q}\AgdaSymbol{)}\AgdaSpace{}%
\AgdaSymbol{(}\AgdaBound{υ}\AgdaSpace{}%
\AgdaOperator{\AgdaInductiveConstructor{▷}}\AgdaSpace{}%
\AgdaBound{R}\AgdaSymbol{))}\AgdaSpace{}%
\AgdaInductiveConstructor{refl}%
\>[79]\AgdaComment{--\ (7)}\<%
\end{code}

Again, the only tricky step is to recognise that case analysis on the
arguments must proceed right-to-left. First we perform case analysis on
\texttt{υ}, and only if \texttt{υ} is a cons do we perform case analysis
on \texttt{τ}, and only if both are conses do we perform case analysis
on \texttt{σ}. For instance, in (6) the two sides simplify to

\begin{verbatim}
 ((σ ⨾ τ) ⨾ (υ ▷ R)) ≡ (σ ⨾ (τ ⨾ (υ ▷ R)))
\end{verbatim}

and the equation follows by the induction hypothesis on \texttt{σ},
\texttt{τ}, and \texttt{υ\ ▷\ R}.

We direct Agda to apply associativity as a rewrite.

\begin{code}%
\>[0]\AgdaSymbol{\{-\#}\AgdaSpace{}%
\AgdaKeyword{REWRITE}\AgdaSpace{}%
\AgdaFunction{assoc}\AgdaSpace{}%
\AgdaSymbol{\#-\}}\<%
\end{code}

\hypertarget{applications}{%
\section{Applications}\label{applications}}

\hypertarget{special-cases-of-substitution}{%
\subsection{Special cases of
substitution}\label{special-cases-of-substitution}}

We define three special cases of substitution.

Substitute for the last variable in the environment (de Bruijn index
zero).

\begin{code}%
\>[0]\AgdaOperator{\AgdaFunction{\AgdaUnderscore{}[\AgdaUnderscore{}]₀}}\AgdaSpace{}%
\AgdaSymbol{:}\<%
\\
\>[0][@{}l@{\AgdaIndent{0}}]%
\>[4]\AgdaSymbol{(}\AgdaBound{N}\AgdaSpace{}%
\AgdaSymbol{:}\AgdaSpace{}%
\AgdaGeneralizable{Γ}\AgdaSpace{}%
\AgdaOperator{\AgdaInductiveConstructor{▷}}\AgdaSpace{}%
\AgdaGeneralizable{A}\AgdaSpace{}%
\AgdaOperator{\AgdaDatatype{⊢}}\AgdaSpace{}%
\AgdaGeneralizable{B}\AgdaSymbol{)}\<%
\\
\>[4]\AgdaSymbol{(}\AgdaBound{M}\AgdaSpace{}%
\AgdaSymbol{:}\AgdaSpace{}%
\AgdaGeneralizable{Γ}\AgdaSpace{}%
\AgdaOperator{\AgdaDatatype{⊢}}\AgdaSpace{}%
\AgdaGeneralizable{A}\AgdaSymbol{)}\<%
\\
\>[0][@{}l@{\AgdaIndent{0}}]%
\>[2]\AgdaSymbol{→}%
\>[762I]\AgdaComment{\myrule}\<%
\\
\>[762I][@{}l@{\AgdaIndent{0}}]%
\>[5]\AgdaGeneralizable{Γ}\AgdaSpace{}%
\AgdaOperator{\AgdaDatatype{⊢}}\AgdaSpace{}%
\AgdaGeneralizable{B}\<%
\\
\>[0]\AgdaBound{N}\AgdaSpace{}%
\AgdaOperator{\AgdaFunction{[}}\AgdaSpace{}%
\AgdaBound{M}\AgdaSpace{}%
\AgdaOperator{\AgdaFunction{]₀}}\AgdaSpace{}%
\AgdaSymbol{=}\AgdaSpace{}%
\AgdaBound{N}\AgdaSpace{}%
\AgdaOperator{\AgdaFunction{[}}\AgdaSpace{}%
\AgdaInductiveConstructor{id}\AgdaSpace{}%
\AgdaOperator{\AgdaInductiveConstructor{▷}}\AgdaSpace{}%
\AgdaBound{M}\AgdaSpace{}%
\AgdaOperator{\AgdaFunction{]}}\<%
\end{code}

This is exactly what we need for beta reduction.

Substitute for the last but one variable in the environment (de Bruijn
index one).

\begin{code}%
\>[0]\AgdaOperator{\AgdaFunction{\AgdaUnderscore{}[\AgdaUnderscore{}]₁}}\AgdaSpace{}%
\AgdaSymbol{:}\<%
\\
\>[0][@{}l@{\AgdaIndent{0}}]%
\>[4]\AgdaSymbol{(}\AgdaBound{N}\AgdaSpace{}%
\AgdaSymbol{:}\AgdaSpace{}%
\AgdaGeneralizable{Γ}\AgdaSpace{}%
\AgdaOperator{\AgdaInductiveConstructor{▷}}\AgdaSpace{}%
\AgdaGeneralizable{A}\AgdaSpace{}%
\AgdaOperator{\AgdaInductiveConstructor{▷}}\AgdaSpace{}%
\AgdaGeneralizable{B}\AgdaSpace{}%
\AgdaOperator{\AgdaDatatype{⊢}}\AgdaSpace{}%
\AgdaGeneralizable{C}\AgdaSymbol{)}\<%
\\
\>[4]\AgdaSymbol{(}\AgdaBound{M}\AgdaSpace{}%
\AgdaSymbol{:}\AgdaSpace{}%
\AgdaGeneralizable{Γ}\AgdaSpace{}%
\AgdaOperator{\AgdaDatatype{⊢}}\AgdaSpace{}%
\AgdaGeneralizable{A}\AgdaSymbol{)}\<%
\\
\>[0][@{}l@{\AgdaIndent{0}}]%
\>[2]\AgdaSymbol{→}%
\>[788I]\AgdaComment{\myrule}\<%
\\
\>[788I][@{}l@{\AgdaIndent{0}}]%
\>[5]\AgdaGeneralizable{Γ}\AgdaSpace{}%
\AgdaOperator{\AgdaInductiveConstructor{▷}}\AgdaSpace{}%
\AgdaGeneralizable{B}\AgdaSpace{}%
\AgdaOperator{\AgdaDatatype{⊢}}\AgdaSpace{}%
\AgdaGeneralizable{C}\<%
\\
\>[0]\AgdaBound{N}\AgdaSpace{}%
\AgdaOperator{\AgdaFunction{[}}\AgdaSpace{}%
\AgdaBound{M}\AgdaSpace{}%
\AgdaOperator{\AgdaFunction{]₁}}\AgdaSpace{}%
\AgdaSymbol{=}\AgdaSpace{}%
\AgdaBound{N}\AgdaSpace{}%
\AgdaOperator{\AgdaFunction{[}}\AgdaSpace{}%
\AgdaSymbol{(}\AgdaInductiveConstructor{id}\AgdaSpace{}%
\AgdaOperator{\AgdaInductiveConstructor{▷}}\AgdaSpace{}%
\AgdaBound{M}\AgdaSymbol{)}\AgdaSpace{}%
\AgdaOperator{\AgdaInductiveConstructor{↑}}\AgdaSpace{}%
\AgdaOperator{\AgdaInductiveConstructor{▷}}\AgdaSpace{}%
\AgdaInductiveConstructor{●}\AgdaSpace{}%
\AgdaOperator{\AgdaFunction{]}}\<%
\end{code}

\hypertarget{an-example}{%
\subsection{An example}\label{an-example}}

The first equation given in the introduction holds trivially.

\begin{code}%
\>[0]\AgdaFunction{introduction}\AgdaSpace{}%
\AgdaSymbol{:}\AgdaSpace{}%
\AgdaSymbol{(}\AgdaGeneralizable{N}\AgdaSpace{}%
\AgdaOperator{\AgdaInductiveConstructor{↑}}\AgdaSymbol{)}\AgdaSpace{}%
\AgdaOperator{\AgdaFunction{[}}\AgdaSpace{}%
\AgdaGeneralizable{M}\AgdaSpace{}%
\AgdaOperator{\AgdaFunction{]₀}}\AgdaSpace{}%
\AgdaOperator{\AgdaDatatype{≡}}\AgdaSpace{}%
\AgdaGeneralizable{N}\<%
\\
\>[0]\AgdaFunction{introduction}\AgdaSpace{}%
\AgdaSymbol{=}\AgdaSpace{}%
\AgdaInductiveConstructor{refl}\<%
\end{code}

It is straightforward to see that the left-hand side simplifies to the
right-hand side. Recall that this took nearly a hundred lines to prove
in the formulation that we used previously!

\hypertarget{a-challenging-exercise}{%
\subsection{A challenging exercise}\label{a-challenging-exercise}}

The following exercise appears in PLFA. It is marked ``stretch'' meaning
it is intended to be challenging.

\begin{code}%
\>[0]\AgdaFunction{double-subst}\AgdaSpace{}%
\AgdaSymbol{:}\AgdaSpace{}%
\AgdaGeneralizable{N}\AgdaSpace{}%
\AgdaOperator{\AgdaFunction{[}}\AgdaSpace{}%
\AgdaGeneralizable{M}\AgdaSpace{}%
\AgdaOperator{\AgdaFunction{]₁}}\AgdaSpace{}%
\AgdaOperator{\AgdaFunction{[}}\AgdaSpace{}%
\AgdaGeneralizable{L}\AgdaSpace{}%
\AgdaOperator{\AgdaFunction{]₀}}\AgdaSpace{}%
\AgdaOperator{\AgdaDatatype{≡}}\AgdaSpace{}%
\AgdaGeneralizable{N}\AgdaSpace{}%
\AgdaOperator{\AgdaFunction{[}}\AgdaSpace{}%
\AgdaGeneralizable{L}\AgdaSpace{}%
\AgdaOperator{\AgdaInductiveConstructor{↑}}\AgdaSpace{}%
\AgdaOperator{\AgdaFunction{]₀}}\AgdaSpace{}%
\AgdaOperator{\AgdaFunction{[}}\AgdaSpace{}%
\AgdaGeneralizable{M}\AgdaSpace{}%
\AgdaOperator{\AgdaFunction{]₀}}\<%
\end{code}

PLFA uses a very different formulation of substitution than the one
given here, and under that formulation the exercise appears quite
challenging---indeed, so far as I know, no one has solved it!

However, with the formulation given here, the exercise becomes trivial.

\begin{code}%
\>[0]\AgdaFunction{double-subst}\AgdaSpace{}%
\AgdaSymbol{=}\AgdaSpace{}%
\AgdaInductiveConstructor{refl}\<%
\end{code}

Both sides simplify by an automatic rewrite with \texttt{lemma-⨟} and
then normalising the compositions yields identical terms.

\hypertarget{a-second-challenge}{%
\subsection{A second challenge}\label{a-second-challenge}}

The following result is the culmination of Chapter Substitution of PLFA.

\begin{code}%
\>[0]\AgdaFunction{commute-subst}\AgdaSpace{}%
\AgdaSymbol{:}\AgdaSpace{}%
\AgdaGeneralizable{N}\AgdaSpace{}%
\AgdaOperator{\AgdaFunction{[}}\AgdaSpace{}%
\AgdaGeneralizable{M}\AgdaSpace{}%
\AgdaOperator{\AgdaFunction{]₀}}\AgdaSpace{}%
\AgdaOperator{\AgdaFunction{[}}\AgdaSpace{}%
\AgdaGeneralizable{L}\AgdaSpace{}%
\AgdaOperator{\AgdaFunction{]₀}}\AgdaSpace{}%
\AgdaOperator{\AgdaDatatype{≡}}\AgdaSpace{}%
\AgdaGeneralizable{N}\AgdaSpace{}%
\AgdaOperator{\AgdaFunction{[}}\AgdaSpace{}%
\AgdaGeneralizable{L}\AgdaSpace{}%
\AgdaOperator{\AgdaFunction{]₁}}\AgdaSpace{}%
\AgdaOperator{\AgdaFunction{[}}\AgdaSpace{}%
\AgdaGeneralizable{M}\AgdaSpace{}%
\AgdaOperator{\AgdaFunction{[}}\AgdaSpace{}%
\AgdaGeneralizable{L}\AgdaSpace{}%
\AgdaOperator{\AgdaFunction{]₀}}\AgdaSpace{}%
\AgdaOperator{\AgdaFunction{]₀}}\<%
\end{code}

In effect, the entire chapter is devoted to proving it. A theory similar
to that of ACCL is developed at length, requiring a few hundred lines of
Agda. Even once the theory is developed, the key lemma,
\texttt{subst-commute}, requires a chain of thirteen equations to prove,
eleven of which required justification.

However, with the formulation given here, the result becomes trivial.

\begin{code}%
\>[0]\AgdaFunction{commute-subst}\AgdaSpace{}%
\AgdaSymbol{=}\AgdaSpace{}%
\AgdaInductiveConstructor{refl}\<%
\end{code}

Both sides simplify by an automatic rewrite with \texttt{lemma-⨟} and
then normalising the compositions yields identical terms.

\hypertarget{conclusion}{%
\section{Conclusion}\label{conclusion}}

A drawback of this technique is that it distinguishes terms that in
traditional notation must be equivalent, such as the terms
\texttt{●\ ↑\ ↑\ ·\ ●\ ↑\ ·\ ●} and \texttt{(●\ ↑\ ·\ ●)\ ↑\ ·\ ●}
mentioned previously. As a result, identifying when terms are equivalent
may become more difficult. On the other hand, if equivalence is
important we are often concerned with normal forms (such as \texttt{β}
and \texttt{η} normal forms) and in that case pushing \texttt{↑} to the
inside as part of normalisation would cause the second term to normalise
to the first, removing the problem of determining equivalence.

Further experience is required. Is explicit weakening useful in
practice? Time will tell.

\bibliographystyle{eptcs}
\bibliography{references}
\end{document}